\documentclass[11pt]{article}
\usepackage{amsmath, afterpage, pdflscape, changepage, multirow, multicol, cleveref, hhline, array, xcolor, verbatim, graphicx, url, fancyhdr, etoolbox, bigstrut, cite, setspace, float, xr, amssymb, mathtools, capt-of, amsthm}

\usepackage[font=small,labelfont=bf]{caption}
\usepackage[hang]{footmisc}
\usepackage[USenglish]{babel}
\usepackage[USenglish]{isodate}
\usepackage[hmargin=2cm,top=1cm,bottom=1.5cm]{geometry}
\usepackage[inline]{enumitem}

\AtBeginEnvironment{tabular}{\doublespacing}

\Crefname{equation}{Eq}{Eqs}

\newcommand{\RestrictTo}[1]{|_{#1}}
\newcommand{\CC}{\mathbb{C}}
\newcommand{\RR}{\mathbb{R}}
\newcommand*\mean[1]{\overline{#1}}
\newcommand*\mystrut[1]{\vrule width0pt height0pt depth#1\relax}

\let\svthefootnote\thefootnote
\newcommand\blankfootnote[1]{%
  \let\thefootnote\relax\footnotetext{#1}%
  \let\thefootnote\svthefootnote%
}

\newtheorem{theorem}{Theorem}%[section]
\newtheorem{lemma}[theorem]{Lemma}
\newtheorem{proposition}[theorem]{Proposition}

\theoremstyle{remark}
\newtheorem{definition}[theorem]{Definition}

\title{\LARGE A novel mathematical model of heterogeneous cell proliferation}

\author{\normalsize Sean T. Vittadello\footnote{School of Mathematical Sciences, Queensland University of Technology, Brisbane, Queensland, Australia.\label{VMS}}\textsuperscript{,}\footnote{\emph{Current address:} Melbourne Integrative Genomics, School of BioSciences, The University of Melbourne, Parkville, Victoria, Australia.}\textsuperscript{,}\renewcommand*{\thefootnote}{\fnsymbol{footnote}}\footnote[2]{Corresponding author: sean.vittadello@unimelb.edu.au}\renewcommand*{\thefootnote}{\arabic{footnote}}, Scott W. McCue\textsuperscript{\ref{VMS}}, Gency Gunasingh\footnote{The University of Queensland Diamantina Institute, The University of Queensland, Brisbane, Queensland, Australia.\label{GH}}, Nikolas K. Haass\textsuperscript{\ref{GH}}, Matthew J. Simpson\textsuperscript{\ref{VMS}}}

\fancypagestyle{specialfooter}{%
\fancyhf{}

\fancyfoot[R]{\footnotesize \emph{\today}}
}

\begin{document}
\captionsetup[figure]{name={Figure}}
%Begin title page
\makeatletter
\begin{titlepage}
\thispagestyle{specialfooter}
\blankfootnote{\textbf{Key words and phrases}: Delay differential equation, Integro-differential equation, Distributed delay, Asymmetric cell division, Induced switching, Transient dynamics\\[0.1cm]}
\centering
\@title\\
\vspace{1.5cm}
\@author\\
\vspace{1.5cm}
\cleanlookdateon
%\@date\\
\vspace{0mm}
\begin{abstract}
\begin{normalsize}\noindent
We present a novel mathematical model of heterogeneous cell proliferation where the total population consists of a subpopulation of slow-proliferating cells and a subpopulation of fast-proliferating cells. The model incorporates two cellular processes, asymmetric cell division and induced switching between proliferative states, which are important determinants for the heterogeneity of a cell population. As motivation for our model we provide experimental data that illustrate the induced-switching process. Our model consists of a system of two coupled delay differential equations with distributed time delays and the cell densities as functions of time. The distributed delays are bounded and allow for the choice of delay kernel. We analyse the model and prove the non-negativity and boundedness of solutions, the existence and uniqueness of solutions, and the local stability characteristics of the equilibrium points. We find that the parameters for induced switching are bifurcation parameters and therefore determine the long-term behaviour of the model. Numerical simulations illustrate and support the theoretical findings, and demonstrate the primary importance of transient dynamics for understanding the evolution of many experimental cell populations.
\end{normalsize}
\end{abstract}
\end{titlepage}
\makeatother

%End title page

\section{Introduction}\label{intro}
Cell proliferation is the fundamental function of the cell cycle \cite{Matson2017}, which is a complex process regulated by both intracellular signals and the extracellular environment \cite{Zhu2019}. Such complexity necessitates that mathematical models of cell proliferation are often restricted to details that are most pertinent to the experimental situation under consideration. The main requirement is that the model must account for progression through the cell cycle in a manner relevant to the cell population and the surrounding environment. Despite all of the underlying complexity the cell cycle has two basic fates, either progression or arrest \cite{Matson2017}. These two cellular fates form the basis of many mathematical models of cell proliferation in the literature, typically based on exponential growth \cite{Lebowitz1974,Webb1986,Swanson2003,Sarapata2014} or logistic growth \cite{Sherratt1990,Maini2004a,Cai2007,Byrne2009,Scott2013,Sarapata2014}. Exponential growth explicitly accounts for progression only, while logistic growth accounts for progression and density-dependent arrest, which can result from contact inhibition \cite{Pavel2018}.

An important detail of the cell cycle not explicitly accounted for in exponential and logistic growth models is the duration of the cell cycle, which is always nonzero and exhibits considerable variation between different cell types and different extracellular environments \cite{Weber2014,Chao2019,Vittadello2019,Vittadello2020}. From a modelling perspective the cell cycle duration is a positive time delay between two sequential cell proliferation events. There are two main types of models which incorporate time delays: one involves functional differential equations \cite{Mackey1994,Byrne1997,Baker1997,Baker1998,Villasana2003,Getto2016,Getto2019,Cassidy2020}, of which delay differential equations are a specific type; and multi-stage models \cite{Yates2017,Simpson2018,Vittadello2018,Vittadello2019,Gavagnin2019}. Models incorporating time delays are consistent with the kinetics of cell proliferation, and can result in a better qualitative and quantitative fit of the model to experimental data \cite{Baker1998}. The inclusion of a time delay must be based on whether the improved model fit outweighs the increase in model complexity arising from additional parameters and, for functional differential equations, an infinite-dimensional state space. Models with time delays are particularly relevant when the transient dynamics of a cell population are of interest, especially when modelling slow-proliferating cells. We briefly note that age-structured models \cite{Arino1995,Gabriel2012,Billy2014,Clairambault2016,Cassidy2019}, which are related to delay differential equations, provide another approach to incorporating realistic cell cycle durations into models of cell population growth.

In this article we introduce a delay differential equation model for cell proliferation in which the cell population consists of a slow-proliferating subpopulation and a fast-proliferating subpopulation. The cells can switch between the \emph{proliferative states} of slow and fast proliferation through two cellular processes: asymmetric cell division and induced switching of proliferative states by surrounding cells. Our model is motivated by the \emph{proliferative heterogeneity}, with respect to cell cycle duration, of tumours, which are often composed of a large proportion of fast-proliferating cells and a small proportion of slow-proliferating cells which can repopulate the fast-proliferating subpopulation \cite{Perego2018,Vallette2019}. The slow-proliferating subpopulation is sometimes considered to be quiescent, or arrested, however it is possible that this subpopulation is actually in a very-slow-proliferating state \cite{Moore2011,Ahn2017}. Experimental studies have found slow-proliferating cells with cell cycle durations greater than four weeks, whereas the predominant fast-proliferating cells have cell cycle durations of around 48 hours \cite{Roesch2010}.

In the literature there are various mathematical models that consider proliferative heterogeneity. Some models account for one proliferating subpopulation \cite{Cassidy2020}, which may undergo asymmetric division \cite{Arino1989,Greene2015}, while the other subpopulations are quiescent or differentiated. Other models consider subpopulations with different proliferative states without any cells switching between the subpopulations \cite{Jin2018}. Our model, and our mathematical analysis of the model, are novel in several ways: \begin{enumerate*}[label=(\arabic*)]
\item for each subpopulation we model a distribution of cell cycle durations using a distributed delay with an arbitrary delay kernel on a bounded interval, which allows us to freely choose an appropriate proliferative state for each subpopulation;
\item cells can switch between the slow- and fast-proliferating subpopulations through two important processes, either during cell division or induced by surrounding cells;
\item we provide formal proofs of existence, uniqueness, non-negativity, and boundedness of the solutions for our model under appropriate initial conditions;
\item the local stability of all equilibrium points is characterised and bifurcation parameters identified, involving the analysis of an interesting transcendental characteristic equation;
\item numerical simulations are provided which illustrate and support the theoretical results, and demonstrate the importance of considering the transient dynamics of experimental cell populations.
\end{enumerate*}

The remainder of this article is organised as follows. In Section~\ref{sec:Motiv} we discuss the biological and mathematical motivations for our model, which we then present in Section~\ref{sec:Model}. Our main analytical results are in Section~\ref{sec:Results} in the form of three theorems: Theorem~\ref{thrm:NonNeg} for non-negativity and boundedness of solutions, Theorem~\ref{thrm:Exist} for the existence and uniqueness of solutions, and Theorem~\ref{thrm:Stability} for the local stability of the equilibrium points. Some examples of numerical simulations of our model are provided in Section~\ref{sec:Sim}, illustrating the long-term dynamics described in Theorem~\ref{thrm:Stability}, and demonstrating the importance of the transient dynamics. Finally, in Section~\ref{sec:Disc} we summarise our results, discuss the utility of our model to describe experimental scenarios, and note some possibilities for future work.

\section{Model motivation}\label{sec:Motiv}
In this section we discuss the biological and mathematical considerations that motivate the development of our model.
\subsection{Biological considerations}\label{subsec:Biol}
\begin{figure}[ht]
\begin{minipage}[t]{0.3\textwidth}
\vspace{0pt} % Necessary for top alignment
\includegraphics[width=0.8\textwidth]{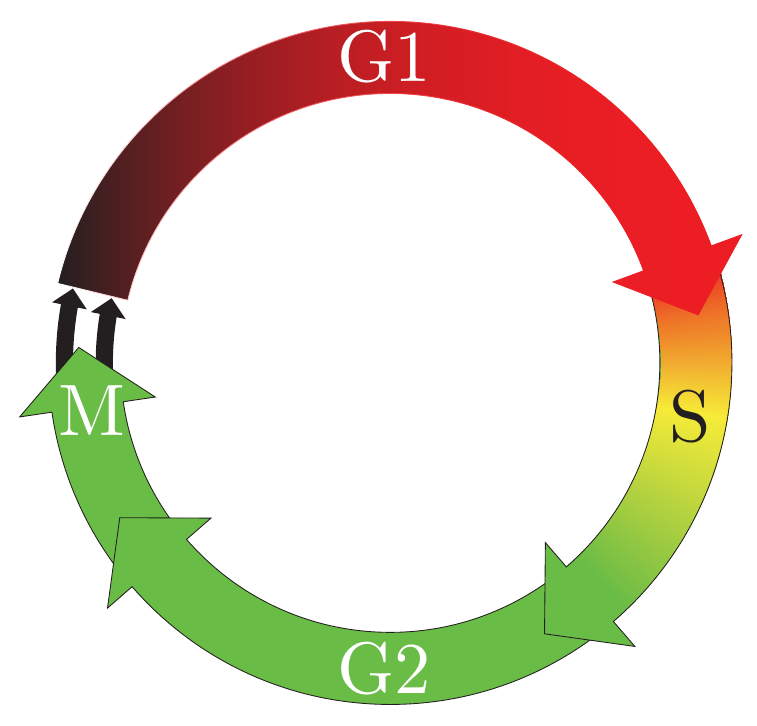}
\end{minipage}\hfill
\begin{minipage}[t]{0.7\textwidth}
\vspace{0pt} % Necessary for top alignment
\caption{Schematic of the eukaryotic cell cycle, indicating the colour of fluorescent ubiquitination-based cell cycle indicator (FUCCI), see \cite{Sakaue_Sawano2008}, in each phase. During very early G1 phase there is no fluorescence as both FUCCI reporters are downregulated. In G1 phase the red FUCCI reporter is upregulated and red fluorescence is observed. During the transition from G1 to S phase, called early S, both the red and green FUCCI reporters are upregulated producing yellow. Through S/G2/M phase the red FUCCI reporter is downregulated and only the green FUCCI reporter is upregulated so that green fluorescence is observed}
\label{fig:Fig1}
\end{minipage}
\end{figure}

The eukaryotic cell cycle (Figure~\ref{fig:Fig1}) is a sequence of four phases, namely gap 1 (G1), synthesis (S), gap 2 (G2) and mitosis (M).
The primary function of the cell cycle is the replication of cellular DNA during S phase, followed by the division of the replicated chromosomes and cytoplasm into two daughter cells during M phase \cite{Vermeulen2003}. Progression through the cell cycle is tightly regulated in normal cells, which are subject to density-dependent contact inhibition producing reversible cell-cycle arrest \cite{McClatchey2012,Puliafito2012}. In cancer cells, however, cell cycle regulation is generally lost \cite{Hanahan2011} resulting in cell populations with proliferative heterogeneity, as exemplified by tumours of solid cancers \cite{Roesch2010,Perego2018}. In particular a small subpopulation of slow-proliferating cells is often present in tumours, and this subpopulation tends to survive anticancer drug treatment and can maintain the tumour by repopulating the fast-proliferating subpopulation \cite{Perego2018,Vallette2019}.

Experimental studies have revealed the highly dynamic nature of intratumoural heterogeneity, which can cause adverse outcomes from cancer therapy, notably drug resistance \cite{Haass2014,Beaumont2016,Ahmed2018,Gallaher2019}. Therefore the nonequilibrium, or transient, state of a tumour tends to be of greater relevance than the equilibrium states. The main purpose of our model is to provide insight into the transient dynamics of intratumoural heterogeneity, specifically with regard to cells switching their proliferative states through cellular mechanisms.

The range of mechanisms leading to proliferative heterogeneity in cancer cell populations are not completely understood, although asymmetric cell division is known to be partly responsible \cite{Bajaj2015,Dey-Guha2015}. Asymmetric cell division is a normal process of stem cell proliferation, required for development and the maintenance of tissue homeostasis, whereby a stem cell divides to produce one daughter stem cell, called \emph{self renewal}, and a second daughter cell that will undergo differentiation. In contrast, symmetric division of a stem cell produces either two daughter stem cells or two daughter cells that will both undergo differentiation \cite{Bajaj2015}. It is known that cancer cells can utilise the pathway of asymmetric cell division to produce heterogeneous populations of cancer cells that support survival of the cancer \cite{Smalley2009,Dey-Guha2011,Dey-Guha2015}.

Another important mechanism contributing to proliferative heterogeneity in cancer cell populations is cell-induced switching between the slow- and fast-proliferating states, which occurs through cell--cell signalling and direct contact between cells \cite{Nelson2002,West2019}. The possibility that cells can switch their proliferative state, either through asymmetric cell division or induced switching by surrounding cells, means that the growth rate of a cell population is highly dependent on the influence of each of these two processes.

\begin{figure}[ht]
\centering
\includegraphics[width=0.7\textwidth]{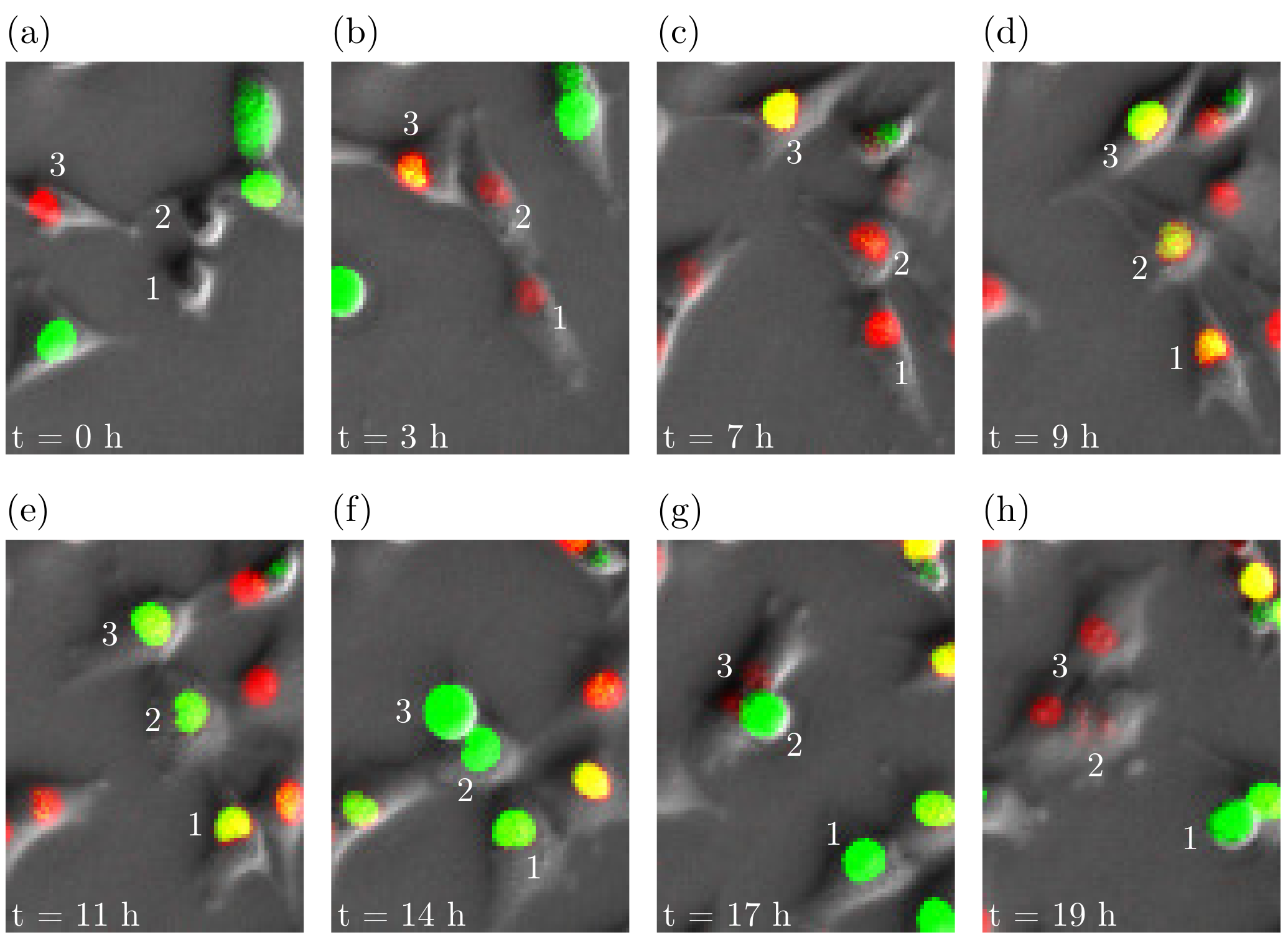}
\caption{Experimental images from a proliferation assay using FUCCI-C8161 melanoma cells, illustrating the possibility of induced switching between proliferative states, as discussed in the text.}
\label{fig:Fig2}
\end{figure}

To illustrate the concept of induced switching we show in Figure~\ref{fig:Fig2} a series of our experimental images from a two-dimensional proliferation assay using FUCCI-C8161 melanoma cells \cite{Haass2014,Spoerri2017}. See Electronic Supplementary Material for further details.
While this experiment is not explicitly designed to study slow- and fast-proliferating subpopulations, it is possible that switching of cell cycle speeds occurs and can be observed by careful inspection of the time-series images. For example, consider the cells labelled 1, 2 and 3 throughout the images. Cells 1 and 2 are daughter cells from the same parent cell and are in an early stage of G1, while cell 3 is a daughter cell from a different parent cell and is in a later stage of G1 at time 0 hours (Figure~\ref{fig:Fig2}(a)). At time 3 hours we observe cell 3 interact closely with cell 2 and not cell 1 (Figure~\ref{fig:Fig2}(b)). At times 7 and 9 hours the three cells continue to progress through the cell cycle with no close interaction between cell 3 and cells 1 and 2 (Figure~\ref{fig:Fig2}(c)--(d)). At 11 hours, cell 3 interacts closely with cell 2 again; cell 2 is in S/G2/M phase, which is further through the cell cycle than cell 1 which is in eS phase (Figure~\ref{fig:Fig2}(e)). At time 14 hours cell 3, which is still close to cell 2, is in M phase and is undergoing mitotic rounding in preparation for cell division (Figure~\ref{fig:Fig2}(f)). At 17 hours, cell 3 undergoes division to produce two daughter cells, cell 2 is undergoing mitotic rounding in preparation for division, whereas cell 1 is in an earlier stage of the cell cycle (Figure~\ref{fig:Fig2}(g)). At time 19 hours cell 2 has divided to produce two daughter cells, whereas cell 1 has only just undergone mitotic rounding in preparation for division (Figure~\ref{fig:Fig2}(h)). These experimental observations illustrate the possibility that cells can switch between states of slow and fast proliferation, induced by surrounding cells. In summary, it seems plausible that cell 2 progresses through the cell cycle faster than cell 1 because of the interactions with cell 3. Indeed, for a given cell line, G1 phase tends to have the most variable duration of the cell cycle phases \cite{Chao2019}, and cells 1 and 2 appear to progress through G1 at a similar rate (Figure~\ref{fig:Fig2}(a)--(c)), so it is possible that asymmetric division does not account for the overall variation in cell cycle duration between cells 1 and 2.

\subsection{Mathematical considerations}\label{subsec:Math}
Delay differential equations are often used when the evolution of the process to be modelled depends on the history of the process, represented as a time delay which may be discrete \cite{Lu1991,Engelborghs2000,Sun2006}, distributed \cite{McCluskey2010,Khasawneh2011,Huang2016,Kaslik2018,Cassidy2020}, or, more generally, state-dependent \cite{Getto2016}. We employ a system of two coupled nonlinear delay differential equations to model the transient dynamics of cell proliferation in a population consisting of slow- and fast-proliferating cells. The time delays are distributed so that only cells of a certain age can proliferate according to an appropriate probability distribution, or \emph{delay kernel}, and our system is therefore of integro-differential type. An alternative to modelling cell proliferation with delay differential equations is to use a multi-stage model, however this is not suitable for modelling the cell proliferation scenario that we consider here. Indeed, cell cycle durations in the multi-stage model are hypoexponentially distributed, and we want to allow for more general distributions of cell cycle duration. Further, the multi-stage model can be difficult to parameterise due to the large number of stages and hence parameters required to represent the cell cycle as stages.

\paragraph{Distributed delays} Standard deterministic mathematical models of cell proliferation, such as exponential and logistic growth models, are based on cell cycle durations with an exponential distribution, which allows for a relatively large probability of arbitrarily small cell cycle durations. Experimental investigations, however, suggest that the duration of the cell cycle, and in particular each cell cycle phase, is not exponentially distributed, rather the hypoexponential distribution is often found to be a reasonable approximation \cite{Weber2014,Yates2017,Vittadello2019,Chao2019,Gavagnin2019}. Our experimental data support these observations (Figure S1, Electronic Supplementary Material). Ordinary differential equations therefore tend to overestimate the cell population growth rate, and may not qualitatively and quantitatively represent the transient growth dynamics of cell populations, particularly for slow-proliferating cells.

The distributions of cell cycle durations for all cell lines are naturally bounded, so unbounded distributions such as the hypoexponential distribution are unrealistic since they theoretically have a nonzero probability of arbitrarily large values. For this reason we consider only bounded distributions for their greater biological realism, and we otherwise allow complete generality for the distributions. Unbounded distributions can be left- or right-truncated to form bounded distributions, and while our model assumes a left bound of zero, alternative left bounds are easily incorporated. Our distributed delays have the following form. Let $X$ denote either the slow-proliferating cells $S$ or fast-proliferating cells $F$, then the distributed delay $\mean{X}(t)$ is defined by
\begin{equation}\label{eq:DistDelay}
\mean{X}(t) = \int_{0}^{U_X} X(t - z)\, g_X (z)\, \mathrm{d}z,
\end{equation}
where the upper limit of integration $U_X \in (0,\infty)$ corresponds to the maximum possible duration of the cell cycle, and $g_X$ is the delay kernel which is normalised so that $\int_{0}^{U_X}\, g_X (z)\, \mathrm{d}z = 1$. The delay kernel is the probability density for the distribution of the cell cycle durations. $\mean{X}(t)$ is a weighted average over the past population densities $X(t-z)$, and corresponds to the subpopulation of cells at time $t$ that are ready to divide.

\paragraph{Contact inhibition} Normal cells are subject to contact inhibition \cite{Pavel2018}, so we assume that the growth rate of each subpopulation at time $t$ has a logistic density dependence given by $\big(1 - P(t)/K\big)$, where $P$ is the total population density and $K$ is the carrying capacity density. For cancer cells, however, contact inhibition may be lost \cite{Hanahan2011} resulting in density-independent growth, so in this case we could set the carrying capacity to infinity in the logistic-growth terms. We can keep the carrying capacity finite in the induced-switching terms, as increasing the cell density beyond a finite carrying capacity would, realistically, increase the probability of cell-induced switching between proliferative states due to surrounding cells.

\paragraph{Proliferation, switching between proliferative states, and apoptosis} Given the cells that are able to undergo division based on the time delays and density constraints, the intrinsic growth rates $0 < r_S \le r_F$ for slow- and fast-proliferating cells, respectively, determine the cells that are parent cells and divide at time $t$. Parent cells can divide symmetrically, where the daughter cells have the same proliferative state as the parent cell, or asymmetrically, where a daughter cell can have a proliferative state different from the parent cell. When a subpopulation of slow-proliferating cells divides to produce twice as many daughter cells, the parameter $\alpha_S$ determines the proportion of these daughter cells that are also slow-proliferating cells, so the proportion $1 - \alpha_S$ of the daughter cells are fast-proliferating cells. Similarly, the parameter $\alpha_F$ determines the proportion of daughter cells from fast-proliferating cells that are also fast-proliferating cells, so the proportion $1 - \alpha_F$ of the daughter cells are slow-proliferating cells. We only consider asymmetric cell division that is self-renewing, that is $\alpha_S$, $\alpha_F \in [\frac{1}{2},1]$, so that the division of a parent cell produces at least one daughter cell with the same proliferative state as the parent, which is the relevant process for cancer cells \cite{Smalley2009,Dey-Guha2011,Dey-Guha2015,Bajaj2015}. While the analysis of our model is valid for $\alpha_S$, $\alpha_F \in [0,1]$, excluding Theorem~\ref{thrm:NonNeg} for non-negativity, our biological motivation necessitates that $\alpha_S$, $\alpha_F \in [\frac{1}{2},1]$.

Induced switching allows for a cell to switch proliferative states at any position of the cell cycle, induced by surrounding cells with a different proliferative state. We make the reasonable assumption that a cell is increasingly likely to switch proliferative states as the density of cells with a different proliferative state increases. This modelling approach would be most realistic when the cell population has a uniform spatial distribution, such as in the proliferation assay in Figure~\ref{fig:Fig2}. The parameter $\beta_S$ corresponds to the per capita interaction strength of fast-proliferating cells to induce slow-proliferating cells to switch to fast proliferation. Similarly, the parameter $\beta_F$ corresponds to the per capita interaction strength of slow-proliferating cells to induce fast-proliferating cells to switch to slow proliferation.

Because we focus on growing tumours, for which proliferation outcompetes cell death, and mechanisms of switching between proliferative states, we simplify our model by not incorporating apoptosis.

\newpage

\section{Mathematical model}\label{sec:Model}
The total cell population consists of the two subpopulations of slow-proliferating cells and fast-proliferating cells, where $P(t) \ge 0$, $S(t) \ge 0$, and $F(t) \ge 0$ are the respective cell densities, so that $P(t) = S(t) + F(t)$. The model is
\begin{align}
\frac{\mathrm{d}S(t)}{\mathrm{d}t} =& \underbrace{\mystrut{10pt} (2\alpha_S - 1) r_S \int_{0}^{U_S} S(t - z)\, g_S (z)\, \mathrm{d}z \strut}_{\substack{\text{slow-proliferating cells from asymmetric} \\ \text{division of slow-proliferating cells}}} \; \underbrace{\mystrut{10pt} \bigg(1 - \frac{\big(S(t) + F(t)\big)}{K}\bigg)}_{\substack{\text{contact inhibition} \\ \text{of proliferation}}} \nonumber \\[0.1cm]
+& \underbrace{\mystrut{10pt} 2(1-\alpha_F) r_F \int_{0}^{U_F} F(t - z)\, g_F (z)\, \mathrm{d}z}_{\substack{\text{slow-proliferating cells from asymmetric} \\ \text{division of fast-proliferating cells}}} \; \underbrace{\mystrut{10pt} \bigg(1 - \frac{\big(S(t) + F(t)\big)}{K}\bigg)}_{\substack{\text{contact inhibition} \\ \text{of proliferation}}} \nonumber \\[0.1cm]
-& \underbrace{\mystrut{10pt} \beta_S S(t) \frac{F(t)}{K}}_{\substack{\text{induced switching} \\ \text{of slow-proliferating cells} \\ \text{to fast-proliferating cells}}} + \underbrace{\mystrut{10pt} \beta_F F(t) \frac{S(t)}{K}}_{\substack{\text{induced switching} \\ \text{of fast-proliferating cells} \\ \text{to slow-proliferating cells}}}, \label{eq:FS1}\\[0.5cm]
\frac{\mathrm{d}F(t)}{\mathrm{d}t} =& \underbrace{\mystrut{10pt} 2(1-\alpha_S) r_S \int_{0}^{U_S} S(t - z)\, g_S (z)\, \mathrm{d}z}_{\substack{\text{fast-proliferating cells from asymmetric} \\ \text{division of slow-proliferating cells}}} \; \underbrace{\mystrut{10pt} \bigg(1 - \frac{\big(S(t) + F(t)\big)}{K}\bigg)}_{\substack{\text{contact inhibition} \\ \text{of proliferation}}} \nonumber \\[0.1cm]
+& \underbrace{\mystrut{10pt} (2\alpha_F - 1) r_F \int_{0}^{U_F} F(t - z)\, g_F (z)\, \mathrm{d}z}_{\substack{\text{fast-proliferating cells from asymmetric} \\ \text{division of fast-proliferating cells}}} \; \underbrace{\mystrut{10pt} \bigg(1 - \frac{\big(S(t) + F(t)\big)}{K}\bigg)}_{\substack{\text{contact inhibition} \\ \text{of proliferation}}} \nonumber \\[0.1cm]
+& \underbrace{\mystrut{10pt} \beta_S S(t) \frac{F(t)}{K}}_{\substack{\text{induced switching} \\ \text{of slow-proliferating cells} \\ \text{to fast-proliferating cells}}} - \underbrace{\mystrut{10pt} \beta_F F(t) \frac{S(t)}{K}}_{\substack{\text{induced switching} \\ \text{of fast-proliferating cells} \\ \text{to slow-proliferating cells}}}, \label{eq:FS2}
\end{align}
where the parameters satisfy
\begingroup
\allowdisplaybreaks
\begin{alignat}{2}
& \text{Intrinsic growth rates:} \hspace{1.5cm} && \text{$r_S$, $r_F \in (0,\infty)$ with $r_S \le r_F$,} \label{eq:param1} \\
& \text{Proportion of symmetric divisions:} \hspace{1.5cm} && \text{$\alpha_S$, $\alpha_F \in [{\textstyle \frac{1}{2}},1]$,} \label{eq:param2} \\
& \text{Maximum cell cycle durations:} \hspace{1.5cm} && \text{$U_S$, $U_F \in (0,\infty)$,} \label{eq:param3} \\
& \text{Induced switching rates:} \hspace{1.5cm} && \text{$\beta_S$, $\beta_F \in [0,\infty)$,} \label{eq:param4}
\end{alignat}
\endgroup
and the mean values of the delay kernels $g_S$ and $g_F$ satisfy
\begin{equation}
\int_{0}^{U_S}\, z\, g_S (z)\, \mathrm{d}z \ge \int_{0}^{U_F}\, z\, g_F (z)\, \mathrm{d}z, \label{eq:meanker}
\end{equation}
so that the mean cell cycle duration for the slow-proliferating cells is not smaller than the mean cell cycle duration for the fast-proliferating cells.

As functional differential equations depend on the solution and perhaps derivatives of the solution at past times it is necessary to specify a function for the initial condition, called the \emph{history function}. Defining $\widehat{U} = \text{max}\{ U_S, U_F \}$, the history function $\phi = (\phi_S,\phi_F)$ for our model satisfies
\begin{align}
&\phi \in C([-\widehat{U},0],\RR_{> 0}^2), \label{eq:history1} \\
&\phi_S + \phi_F \in C([-\widehat{U},0],(0,K)), \label{eq:history2}
\end{align}
where $C([-\widehat{U},0],\RR_{> 0}^2)$ is the space of continuous functions on $[-\widehat{U},0]$ into $\RR_{> 0}^2$ and $C([-\widehat{U},0],(0,K))$ is the space of continuous functions on $[-\widehat{U},0]$ into $(0,K)$. Note that state space is therefore an infinite-dimensional function space. For bounded delays the state space is typically the Banach space $C([-h,0],\RR^n)$, for some $h \in (0,\infty)$, of continuous functions $\chi \colon [-h,0] \to \RR^n$ on the closed interval $[-h,0]$ under the supremum norm $\lVert \cdot \rVert$ defined by $\lVert \chi \rVert = \sup \{\, \lVert \chi(t) \rVert_2 \mid t \in [-h,0] \,\}$, where $\lVert \cdot \rVert_2$ is the Euclidean norm on $\RR^n$. In our case, state space $C$ is defined by
\begin{equation}\label{eq:StateSpace}
C = C([-\widehat{U},0],\RR^2).
\end{equation}

Finally, we note that adding Equations~\eqref{eq:FS1} and \eqref{eq:FS2} gives
\begin{equation}\label{eq:Sum}
\frac{\mathrm{d}P(t)}{\mathrm{d}t} = \bigg(r_S \int_{0}^{U_S} S(t - z)\, g_S (z)\, \mathrm{d}z + r_F \int_{0}^{U_F} F(t - z)\, g_F (z)\, \mathrm{d}z \bigg) \bigg(1 - \frac{P(t)}{K}\bigg),
\end{equation}
so, from the perspective of the whole population, asymmetric division and induced switching have no net effects. Moreover, if we consider the total population as composed of cells in the same proliferative state with $r_S = r_F$, and $g_S (z) = g_F (z) = \delta (z)$ is the Dirac kernel for zero delay, then Equation~\eqref{eq:Sum} reduces to the logistic growth model (Electronic Supplementary Material).

\clearpage

\section{Main results}\label{sec:Results}
We now discuss our analysis of Equations~\eqref{eq:FS1} and \eqref{eq:FS2}, namely non-negativity and boundedness of solutions, existence and uniqueness of solutions, and local stability analysis of the equilibrium points. A solution for Equations~\eqref{eq:FS1} and \eqref{eq:FS2} means the following \cite{Kuang1993,Smith2011}:

\begin{definition}[Solution]
Consider the system of delay differential equations~\eqref{eq:FS1} and \eqref{eq:FS2} with parameters, delay kernels, and history functions that satisfy \eqref{eq:param1}--\eqref{eq:history2}. A \emph{solution} for the system is a function $(S,F) \in C([-\widehat{U},u),\RR_{\ge 0}^2)$ for some $u \in [0,\infty]$ such that:
\begin{itemize}
\item $S$ and $F$ are differentiable on $(0,u)$ and right-differentiable at $0$;
\item $(S,F)$ satisfies Equations~\eqref{eq:FS1} and \eqref{eq:FS2} for $t \in [0,u)$.
\end{itemize}
Additionally, $(S,F)$ is a solution with \emph{initial condition} $\phi = (\phi_S,\phi_F) \in C([-\widehat{U},0],\RR_{> 0}^2)$ if
\begin{itemize}
\item $(S,F) \rvert_{[-\widehat{U},0]} = \phi$.
\end{itemize}
\end{definition}

\subsection{Non-negativity and boundedness}
Since the dependent variables $S(t)$ and $F(t)$ in Equations~\eqref{eq:FS1} and \eqref{eq:FS2} represent cell densities they must assume non-negative values at all times. Further, the densities of normal cells are bounded above by the carrying capacity density $K$ arising from contact inhibition. Cancer cells typically have unregulated growth due to the loss of contact inhibition, however continued growth depends on environmental conditions such as nutrient availability, so it is reasonable to assume that the density of cancer cells is also bounded by a carrying capacity.

\begin{theorem}[Non-negativity and boundedness of solutions]\label{thrm:NonNeg}\hfill \\
Let $(S,F)$ be a solution for the system of delay differential equations~\eqref{eq:FS1} and \eqref{eq:FS2} with parameters, delay kernels, and history functions that satisfy \eqref{eq:param1}--\eqref{eq:history2}. Then $S(t)$, $F(t) \in [0,K]$ for all $t > 0$, therefore the solutions $S$ and $F$ are non-negative and bounded.
\end{theorem}
\noindent
We give an elementary proof of this theorem as it facilitates understanding of the non-negativity and boundedness of the solutions. We first require a lemma.

\begin{lemma}\label{P(t)}
Let $(S,F)$ be a solution for the system of delay differential equations~\eqref{eq:FS1} and \eqref{eq:FS2} with parameters, delay kernels, and history functions that satisfy \eqref{eq:param1}--\eqref{eq:history2}. If there exists $T > 0$ such that the distributed delays satisfy the relations $\mean{S}(t) \ge 0$ and $\mean{F}(t) \ge 0$ for all $t \in (0,T]$ then $P(t) \in [0,K]$ for all $t \in (0,T]$.
\end{lemma}

\begin{proof}
If $P(t_1) > K$ for some $t_1 \in (0,T]$ then, since $P(0) < K$ by Equation~\eqref{eq:history2}, we may assume without loss of generality that $\mathrm{d}P(t) / \mathrm{d}t \, \rvert_{t = t_1} > 0$. Since Equation~\eqref{eq:Sum} gives $\mathrm{d}P(t) / \mathrm{d}t \, \rvert_{t = t_1} \le 0$, we have a contradiction. It follows that $P(t) \le K$ for all $t \in (0,T]$. Similarly, if $P(t_2) < 0$ for some $t_2 \in (0,T]$ then, since $P(0) > 0$ by Equation~\eqref{eq:history2}, we may assume without loss of generality that $\mathrm{d}P(t) / \mathrm{d}t \, \rvert_{t = t_2} < 0$. Since Equation~\eqref{eq:Sum} gives $\mathrm{d}P(t) / \mathrm{d}t \, \rvert_{t = t_2} \ge 0$, we have a contradiction. It follows that $P(t) \ge 0$ for all $t \in (0,T]$.
\end{proof}

\begin{proof}[Theorem~\ref{thrm:NonNeg}]
It suffices to prove that $S(t) \ge 0$ and $F(t) \ge 0$ for all $t > 0$, for then it follows from Lemma~\ref{P(t)} that $S(t) \le K$ and $F(t) \le K$ for all $t > 0$. Define $t_1$ and $t_2$ by $t_1 = \inf \{\, t > 0 \mid S(t) < 0 \,\}$ and $t_2 = \inf \{\, t > 0 \mid F(t) < 0 \,\}$. We consider the infima in the extended real numbers, so that $t_1$, $t_2 \in (0,+\infty]$, where either infimum is equal to $+\infty$ if the corresponding set is empty. The proof consists of four separate cases which are proved similarly. We demonstrate one case here, and provide the complete proof in the Electronic Supplementary Material.

Case 1: Let $\beta_S - \beta_F \ge 0$ and suppose $S(t) < 0$ for some $t > 0$.\\
Note that $t_1 \in \RR$. If $t_1 < t_2$ then choose $t_3 \in (t_1,t_2)$ such that $S(t_3) < 0$, $\mathrm{d}S(t) / \mathrm{d}t \, \rvert_{t = t_3} < 0$, and the delays satisfy $\mean{S}(t_3) \ge 0$ and $\mean{F}(t_3) \ge 0$. Then, since $P(t_3) \le K$ by Lemma~\ref{P(t)} and since $F(t_3) \ge 0$, Equation~\eqref{eq:FS1} gives $\mathrm{d}S(t) / \mathrm{d}t \rvert_{t = t_3} \ge 0$, a contradiction.\\
If $t_1 \ge t_2$ then, since $F(0) > 0$ and $F(t_2) = 0$, there exists $t_3 \in (0,t_2)$ such that $\mathrm{d}F(t) / \mathrm{d}t \rvert_{t = t_3} < 0$. Then, since $P(t_3) \le K$ by Lemma~\ref{P(t)}, since the delays satisfy $\mean{S}(t_3) \ge 0$ and $\mean{F}(t_3) \ge 0$, and since $S(t_3)$, $F(t_3) \ge 0$, Equation~\eqref{eq:FS2} gives $\mathrm{d}F(t) / \mathrm{d}t \rvert_{t = t_3} \ge 0$, a contradiction. We conclude that $S(t) \ge 0$ for all $t > 0$.
\end{proof}

\subsection{Existence and uniqueness}
We begin by introducing some simplifying notation. For $\rho$ in the state space $C$ we denote the component functions by $\rho_S$ and $\rho_F$ so that $\rho = (\rho_S,\rho_F)$, and then define $\mean{\rho_S}$ and $\mean{\rho_F}$ by
\begin{equation}\label{eq:rho}
\mean{\rho_S} = \int_{0}^{U_S} \rho_S (-z)\, g_S (z)\, \mathrm{d}z \quad \text{and} \quad \mean{\rho_F} = \int_{0}^{U_F} \rho_F (-z)\, g_F (z)\, \mathrm{d}z \, .
\end{equation}
Now we define $f \colon C \to \RR^2$ by
\begin{equation}\label{eq:f}
f(\rho) =
\left[
\begin{array}{ll}
\displaystyle \Big((2\alpha_S - 1) r_S \mean{\rho_S} + 2(1-\alpha_F) r_F \mean{\rho_F} \Big) \bigg(1 - \frac{(\rho_S + \rho_F) (0)}{K}\bigg) \\[0.2cm]
\displaystyle \quad -\frac{(\beta_S - \beta_F)}{K} \rho_S (0) \rho_F (0) \\[0.4cm]
\displaystyle \Big(2(1-\alpha_S) r_S \mean{\rho_S} + (2\alpha_F - 1) r_F \mean{\rho_F} \Big) \bigg(1 - \frac{(\rho_S + \rho_F) (0)}{K}\bigg) \\[0.2cm]
\displaystyle \quad +\frac{(\beta_S - \beta_F)}{K} \rho_S (0) \rho_F (0)
\end{array}
\right].
\end{equation}
Note that $f$ is continuous. If $(S,F)$ is a solution for Equations~\eqref{eq:FS1} and \eqref{eq:FS2}, $t \ge 0$, and we define $(S_t,F_t) \in C ([-\widehat{U},0],\RR^2)$ by $(S_t (r),F_t (r)) = (S(t+r),F(t+r))$ for $r \in [-\widehat{U},0]$, then $f((S_t,F_t)) = [\mathrm{d}S(t) / \mathrm{d}t, \mathrm{d}F(t) / \mathrm{d}t]^T$ from \eqref{eq:FS1} and \eqref{eq:FS2}.

\begin{theorem}[Existence and uniqueness of solutions]\label{thrm:Exist}\hfill \\
Consider the system of delay differential equations~\eqref{eq:FS1} and \eqref{eq:FS2} with parameters, delay kernels, and history functions which satisfy \eqref{eq:param1}--\eqref{eq:history2}. Then there exists a unique solution $(S,F) \in C([-\widehat{U},\infty),\RR_{\ge 0}^2)$ of \eqref{eq:FS1} and \eqref{eq:FS2}.
\end{theorem}

\begin{proof}
Here we give an outline of the proof. The complete proof is provided in the Electronic Supplementary Material. We first show that $f$ defined in Equation~\eqref{eq:f} satisfies the following Lipschitz condition on every bounded subset of $C$: for all $M > 0$ there exists $L > 0$ such that for every $\rho$, $\psi \in C([-\widehat{U},0],\RR^2)$ with $\lVert \rho \rVert$, $\lVert \psi \rVert \le M$ we have $\lVert f(\rho) - f(\psi) \rVert_2 \le L \lVert \rho - \psi \rVert$.

To further simplify the notation in Equation~\eqref{eq:f} we define $\kappa_1 = (2\alpha_S - 1) r_S$, $\kappa_2 = 2(1-\alpha_F) r_F$, $\kappa_3 = 2(1-\alpha_S) r_S$, $\kappa_4 = (2\alpha_F - 1) r_F$, and $\kappa_5 = (\beta_S - \beta_F) / K$. Now,
\begingroup
\allowdisplaybreaks
\begin{align*}
f(\rho) - f(\psi) &=
\left[
\begin{array}{ll}
\displaystyle \big(\kappa_1 ( \mean{\rho_S} - \mean{\psi_S} ) + \kappa_2 ( \mean{\rho_F} - \mean{\psi_F} ) \big) \bigg(1 - \frac{(\rho_S + \rho_F) (0)}{K}\bigg) \\[0.3cm]
\displaystyle + \Big(\kappa_1 \mean{\psi_S} + \kappa_2 \mean{\psi_F} \Big) \bigg(\frac{(\psi_S - \rho_S) (0) + (\psi_F - \rho_F) (0)}{K}\bigg) \\[0.3cm]
+ \kappa_5 (\psi_S - \rho_S) (0) \psi_F (0) + \kappa_5 (\psi_F - \rho_F) (0) \rho_S (0) \\[0.4cm]
\displaystyle \big(\kappa_3 ( \mean{\rho_S} - \mean{\psi_S} ) + \kappa_4 ( \mean{\rho_F} - \mean{\psi_F} ) \big) \bigg(1 - \frac{(\rho_S + \rho_F) (0)}{K}\bigg) \\[0.3cm]
\displaystyle + \Big(\kappa_3 \mean{\psi_S} + \kappa_4 \mean{\psi_F} \Big) \bigg(\frac{(\psi_S - \rho_S) (0) + (\psi_F - \rho_F) (0)}{K}\bigg) \\[0.3cm]
+ \kappa_5 (\rho_S - \psi_S)(0) \psi_F (0) + \kappa_5 ( \rho_F - \psi_F)(0) \rho_S (0)
\end{array}
\right]
\end{align*}
\endgroup
so, using the triangle inequality, we obtain
\begingroup
\allowdisplaybreaks
\begin{align*}
\lVert f(\rho) - f(\psi) \rVert_2 
\le \Bigg(\bigg(\sqrt{\kappa_1^2 + \kappa_3^2} + \sqrt{\kappa_2^2 + \kappa_4^2}\bigg) \bigg(1 + \frac{4M}{K} \bigg) 
+ 2 \sqrt{2} \lvert \kappa_5 \rvert M \Bigg) \lVert \rho - \psi \rVert,
\end{align*}
so we can set $L$ to be
\begin{equation*}
L = \bigg(\sqrt{\kappa_1^2 + \kappa_3^2} + \sqrt{\kappa_2^2 + \kappa_4^2}\bigg) \bigg(1 + \frac{4M}{K} \bigg) 
+ 2 \sqrt{2} \lvert \kappa_5 \rvert M
\end{equation*}
\endgroup
and then $f$ satisfies the Lipschitz condition. Then \cite[Page 32, Theorem 3.7]{Smith2011} provides local existence and uniqueness of solutions for the system \eqref{eq:FS1} and \eqref{eq:FS2}. Since our solutions of interest are bounded by Theorem~\ref{thrm:NonNeg}, it follows from \cite[Page 37, Proposition 3.10]{Smith2011} that the solutions are continuable to all positive time.
\end{proof}

\subsection{Local stability}
Here we consider the local stability analysis of the equilibrium points for the system in \eqref{eq:FS1} and \eqref{eq:FS2}, and show that $\beta_S$ and $\beta_F$ are bifurcation parameters with bifurcation point when $\beta_S = \beta_F$. We will prove the following theorem.

\begin{theorem}[Local stability]\label{thrm:Stability}\hfill \\
Consider the system of delay differential equations~\eqref{eq:FS1} and \eqref{eq:FS2} with parameters, delay kernels, and history functions that satisfy \eqref{eq:param1}--\eqref{eq:history2}.
\begin{itemize}[label=$\bullet$]
\item When $\beta_S \ne \beta_F$ the system has the three equilibrium points $(0,0)$, $(0,K)$, and $(K,0)$ with the following properties:
	\begin{itemize}
	\item $(0,0)$ is locally unstable.
	\item If $\beta_S > \beta_F$ then $(K,0)$ is locally unstable and $(0,K)$ is locally stable.
	\item If $\beta_S < \beta_F$ then $(K,0)$ is locally stable and $(0,K)$ is locally unstable.
	\end{itemize}
\item When $\beta_S = \beta_F$ the system has infinitely many equilibrium points corresponding to the line segment joining $(K,0)$ and $(0,K)$, all of which are locally stable.
\end{itemize}
The parameters $\beta_S$ and $\beta_F$ are therefore bifurcation parameters.
\end{theorem}

\noindent
Note that, since equilibrium points for delay differential equations are functions in a Banach space, we have $(0,0)$, $(0,K)$, $(K,0) \in C([-\widehat{U},\infty),\RR_{\ge 0}^2)$.

The proof of Theorem~\ref{thrm:Stability} follows immediately from Propositions~\ref{prop:Equil0}, \ref{prop:Equil1}, \ref{prop:Equil2} and \ref{prop:Equil3}. We begin by non-dimensionalising Equations~\eqref{eq:FS1} and \eqref{eq:FS2}, and then linearising the non-dimensional system about the equilibrium points. Denoting the dimensionless variables with a caret, we define $\hat{t} = r_F t$, $\hat{S} (\hat{t}) = S(t) / K$ and $\hat{F} (\hat{t}) = F(t) / K$. We also define the dimensionless parameters $r = r_S / r_F$ and $\beta = (\beta_S - \beta_F) / r_F$. Equations~\eqref{eq:FS1} and \eqref{eq:FS2} then become, dropping the caret notation for simplicity,
\begin{align}
\frac{\mathrm{d} S(t)}{\mathrm{d} t} &= \bigg((2\alpha_S - 1) r \int_{0}^{U_S} S(t - r_F z)\, g_S (z)\, \mathrm{d}z + 2(1-\alpha_F) \int_{0}^{U_F} F(t - r_F z)\, g_F (z)\, \mathrm{d}z \bigg) \nonumber \\[0.2cm]
& \quad \times \big(1 - S(t) - F(t)\big) -\beta S(t) F(t), \label{eq:FS5}\\[0.3cm]
\frac{\mathrm{d} F(t)}{\mathrm{d} t} &= \bigg(2(1-\alpha_S) r \int_{0}^{U_S} S(t - r_F z)\, g_S (z)\, \mathrm{d}z + (2\alpha_F - 1) \int_{0}^{U_F} F(t - r_F z)\, g_F (z)\, \mathrm{d}z \bigg)  \nonumber \\[0.2cm]
& \quad \times \big(1 - S(t) - F(t)\big) +\beta S(t) F(t). \label{eq:FS6}
\end{align}
Since $S$ and $F$ are cell densities, hence non-negative, we only consider equilibrium points $(S^{\text{*}},F^{\text{*}}) \in C([-\widehat{U},\infty),\RR_{\ge 0}^2)$. To find the equilibrium points we substitute $S = S^{\text{*}}$ and $F = F^{\text{*}}$ into \eqref{eq:FS5} and \eqref{eq:FS6} to give
\begin{align}
0 &= \big((2\alpha_S - 1) r S^{\text{*}} + 2(1-\alpha_F) F^{\text{*}}\big) \big(1 - S^{\text{*}} - F^{\text{*}}\big) -\beta S^{\text{*}} F^{\text{*}}, \label{eq:FS7}\\[0.0cm]
0 &= \big(2(1-\alpha_S) r S^{\text{*}} + (2\alpha_F - 1) F^{\text{*}}\big) \big(1 - S^{\text{*}} - F^{\text{*}}\big) +\beta S^{\text{*}} F^{\text{*}}, \label{eq:FS8}
\end{align}
hence $(S^{\text{*}},F^{\text{*}}) = (0,0)$, $(1,0)$ or $(0,1)$ when $\beta \ne 0$. When $\beta = 0$ the equilibrium points consist of the two lines $(S^{\text{*}},F^{\text{*}}) = (u,1 - u)$ for all $u \in \RR$ and $(S^{\text{*}},F^{\text{*}}) = (u,-r u)$ for all $u \in \RR$, for which the non-negative points are $(S^{\text{*}},F^{\text{*}}) = (u,1 - u)$ for all $u \in [0,1]$ and $(S^{\text{*}},F^{\text{*}}) = (0,0)$.

To examine the local stability of the equilibrium points $(S^{\text{*}},F^{\text{*}})$ we linearise the system in Equations~\eqref{eq:FS5} and \eqref{eq:FS6} about each point. Defining $x(t) = S(t) - S^{\text{*}}$ and $y(t) = F(t) - F^{\text{*}}$ we obtain the linearised system:
\begin{align}
\frac{\mathrm{d} x(t)}{\mathrm{d} t} &=
\bigg((2\alpha_S - 1) r \Big(\int_{0}^{U_S} x(t - r_F z)\, g_S (z)\, \mathrm{d}z + S^{\text{*}}\Big) \nonumber \\
& \qquad + 2(1-\alpha_F) \Big(\int_{0}^{U_F} y(t - r_F z)\, g_F (z)\, \mathrm{d}z + F^{\text{*}} \Big) \bigg) \nonumber \\[0.0cm]
& \qquad \times \big( 1 - x(t) - y(t) - S^{\text{*}} - F^{\text{*}} \big) - \beta \big( x(t) + S^{\text{*}}\big) \big(y(t) + F^{\text{*}}\big) \nonumber \\[0.0cm]
& \sim \bigg((2\alpha_S - 1) r \int_{0}^{U_S} x(t - r_F z)\, g_S (z)\, \mathrm{d}z + 2(1-\alpha_F) \int_{0}^{U_F} y(t - r_F z)\, g_F (z)\, \mathrm{d}z \bigg) \nonumber \\[0.0cm]
& \qquad \times \big(1 - S^{\text{*}} - F^{\text{*}} \big) \nonumber \\[0.0cm]
& \qquad + \Big((2\alpha_S - 1) r S^{\text{*}} + 2(1-\alpha_F) F^{\text{*}} \Big) \big(1 - x(t) - y(t) - S^{\text{*}} - F^{\text{*}} \big) \nonumber \\[0.0cm]
& \qquad - \beta \big( x(t) F^{\text{*}} + y(t) S^{\text{*}} + S^{\text{*}}F^{\text{*}} \big), \quad \text{as $x(t)$, $y(t) \to 0$}, \label{eq:FS9}
\end{align}

\vspace{-1.5cm}

\begin{align}
\frac{\mathrm{d} y(t)}{\mathrm{d} t} &=
\bigg(2(1-\alpha_S) r \Big(\int_{0}^{U_S} x(t - r_F z)\, g_S (z)\, \mathrm{d}z + S^{\text{*}}\Big) \nonumber \\
& \qquad + (2\alpha_F - 1) \Big(\int_{0}^{U_F} y(t - r_F z)\, g_F (z)\, \mathrm{d}z + F^{\text{*}} \Big) \bigg)  \nonumber \\[0.0cm]
& \qquad \times \big(1 - x(t) - y(t) - S^{\text{*}} - F^{\text{*}} \big) + \beta \big(x(t) + S^{\text{*}}\big) \big(y(t) + F^{\text{*}}\big) \nonumber \\[0.0cm]
& \sim \bigg(2(1-\alpha_S) r \int_{0}^{U_S} x(t - r_F z)\, g_S (z)\, \mathrm{d}z + (2\alpha_F - 1) \int_{0}^{U_F} y(t - r_F z)\, g_F (z)\, \mathrm{d}z \bigg) \nonumber \\[0.0cm]
& \qquad \times \big(1 - S^{\text{*}} - F^{\text{*}} \big) \nonumber \\[0.0cm]
& \qquad + \Big(2(1-\alpha_S) r S^{\text{*}} + (2\alpha_F - 1) F^{\text{*}} \Big) \big(1 - x(t) - y(t) - S^{\text{*}} - F^{\text{*}} \big) \nonumber \\[0.0cm]
& \qquad + \beta \big( x(t) F^{\text{*}} + y(t) S^{\text{*}} + S^{\text{*}}F^{\text{*}} \big),  \quad \text{as $x(t)$, $y(t) \to 0$}. \label{eq:FS10}
\end{align}
By the Principle of Linearised Stability \cite[Page 240, Theorem 6.8]{Diekmann1995} it suffices to consider the stability of the equilibrium points for the linearisation in Equations~\eqref{eq:FS9} and \eqref{eq:FS10}.

\begin{proposition}[Equilibrium point $(0,0)$]\label{prop:Equil0}\hfill \\
Consider the system of delay differential equations~\eqref{eq:FS1} and \eqref{eq:FS2} with parameters, delay kernels, and history functions which satisfy \eqref{eq:param1}--\eqref{eq:history2}. Then $(0,0)$ is locally unstable.
\end{proposition}

Proposition~\ref{prop:Equil0} follows immediately from Proposition~\ref{prop:Transcendental}, in which we analyse the transcendental characteristic equation associated with $(0,0)$ of the linearised system \eqref{eq:FS9} and \eqref{eq:FS10} to show that the characteristic equation has at least one zero in $\CC$ with positive real part. While there are alternative methods for proving Proposition~\ref{prop:Equil0} (Electronic Supplementary Material), we consider the direct approach of analysing the associated transcendental characteristic equation to have mathematical relevance for other studies involving delay differential equations. Indeed, transcendental characteristic equations are generally difficult to analyse \cite[Chapter XI]{Diekmann1995}, so new analysis of such equations is of mathematical interest.

For $(S^{\text{*}},F^{\text{*}}) = (0,0)$, Equations~\eqref{eq:FS9} and \eqref{eq:FS10} become
\begin{align}
&\frac{\mathrm{d} x(t)}{\mathrm{d} t} = 
(2\alpha_S - 1) r \int_{0}^{U_S} x(t - r_F z)\, g_S (z)\, \mathrm{d}z + 2(1-\alpha_F) \int_{0}^{U_F} y(t - r_F z)\, g_F (z)\, \mathrm{d}z, \label{eq:FS11} \\[0.0cm]
&\frac{\mathrm{d} y(t)}{\mathrm{d} t} = 
2(1-\alpha_S) r \int_{0}^{U_S} x(t - r_F z)\, g_S (z)\, \mathrm{d}z + (2\alpha_F - 1) \int_{0}^{U_F} y(t - r_F z)\, g_F (z)\, \mathrm{d}z. \label{eq:FS12}
\end{align}
Equations~\eqref{eq:FS11} and \eqref{eq:FS12} have a solution of the form
\begin{equation}\label{eq:solution}
\begin{bmatrix} x(t)\\ y(t) \end{bmatrix} = \begin{bmatrix} c_1\\ c_2 \end{bmatrix} e^{\lambda t}, \quad \text{where} \; \begin{bmatrix} c_1\\ c_2 \end{bmatrix} \in \CC^2 \; \text{is nonzero and} \; \lambda \in \CC,
\end{equation}
so substitution gives
\begin{align}
&e^{\lambda t} \begin{bmatrix} \lambda & 0 \\ 0 & \lambda \end{bmatrix} \begin{bmatrix} c_1\\ c_2 \end{bmatrix} = \nonumber \\[0.cm]
&\begin{bmatrix}
\displaystyle (2\alpha_S - 1) r \int_{0}^{U_S} e^{-\lambda r_F z} \, g_S (z) \, \mathrm{d}z & \quad 2(1-\alpha_F) \int_{0}^{U_F} e^{-\lambda r_F z} \, g_F (z) \, \mathrm{d}z \\[0.5cm]
\displaystyle 2(1-\alpha_S) r \int_{0}^{U_S} e^{-\lambda r_F z} \, g_S (z)\, \mathrm{d}z & (2\alpha_F - 1) \int_{0}^{U_F} e^{-\lambda r_F z} \, g_F (z)\, \mathrm{d}z
\end{bmatrix}
\begin{bmatrix} c_1\\ c_2 \end{bmatrix} e^{\lambda t}. \label{eq:FS13}
\end{align}
To ensure that $(c_1 , c_2 )^{\intercal} \neq 0$ we must have the characteristic equation
\begin{align}
G(\lambda) &= \begin{vmatrix}
\displaystyle (2\alpha_S - 1) r \int_{0}^{U_S} e^{-\lambda r_F z} \, g_S (z) \, \mathrm{d}z \, - \lambda & \quad 2(1-\alpha_F) \int_{0}^{U_F} e^{-\lambda r_F z} \, g_F (z) \, \mathrm{d}z \\[0.2cm]
\displaystyle 2(1-\alpha_S) r \int_{0}^{U_S} e^{-\lambda r_F z} \, g_S (z)\, \mathrm{d}z & (2\alpha_F - 1) \int_{0}^{U_F} e^{-\lambda r_F z} \, g_F (z)\, \mathrm{d}z \, - \lambda
\end{vmatrix} \nonumber \\[0.2cm]
&= \lambda^2 - \lambda \bigg((2\alpha_S - 1) r \int_{0}^{U_S} e^{-\lambda r_F z} \, g_S (z) \, \mathrm{d}z + (2\alpha_F - 1) \int_{0}^{U_F} e^{-\lambda r_F z} \, g_F (z)\, \mathrm{d}z \bigg) \nonumber \\[0.2cm]
& \hspace{0.8cm} + (2\alpha_S + 2\alpha_F - 3) r \int_{0}^{U_S} \int_{0}^{U_F} e^{-\lambda r_F (z+v)} \, g_F (v) \, g_S (z) \, \mathrm{d}v \, \mathrm{d}z \label{eq:Transcendental}
\end{align}
equal to zero. The zeros of the transcendental equation $G(\lambda)$ are the eigenvalues. Proposition~\ref{prop:Transcendental} shows that $G(\lambda)$ has at least one zero with positive real part, so the equilibrium point $(0,0)$ is locally unstable. In the proof of Proposition~\ref{prop:Transcendental} we consider three cases for $G(\lambda)$ in Equation~\eqref{eq:Transcendental}, depending on whether $2\alpha_S + 2\alpha_F - 3$ is negative, zero, or positive. To understand the physical interpretation of $2\alpha_S + 2\alpha_F - 3$, first note that $2\alpha_S + 2\alpha_F - 3 = (2\alpha_S - 1) - 2(1-\alpha_F)$. Referring to Equation~\eqref{eq:FS5}, $(2\alpha_S - 1) - 2(1-\alpha_F)$ is the difference between the proportion of slow-proliferating parent cells that produce slow-proliferating daughter cells beyond self renewal, and the proportion of fast-proliferating parent cells that produce slow-proliferating daughter cells, at a given time. A similar interpretation follows by referring to Equation~\eqref{eq:FS6} and noting that $2\alpha_S + 2\alpha_F - 3 = (2\alpha_F - 1) - 2(1-\alpha_S)$. When $2\alpha_S + 2\alpha_F - 3$ is negative or zero we use the intermediate value theorem to prove that $G(\lambda)$ has a zero in $\RR$ which is positive. When $2\alpha_S + 2\alpha_F - 3$ is positive, however, we require a different approach involving Cauchy's argument principle, which we now outline.

Let $\Omega$ be a non-empty connected open set, let $\Gamma$ be a closed curve in $\Omega$ with positive, or counter-clockwise, orientation which is homologous to zero with respect to $\Omega$, and let $h$ be a meromorphic function on $\Omega$ with no zeros or poles on $\Gamma$. Then Cauchy's argument principle is \cite[Page 152, Theorem 18]{Ahlfors1979}
\begin{equation}\label{eq:Cauchy}
\frac{1}{2\pi i} \oint_{\Gamma} \frac{h^{\prime}(\lambda)}{h(\lambda)}\,\mathrm{d}\lambda = \mathcal{Z} - \mathcal{P},
\end{equation}
where $\mathcal{Z}$ is the number of zeros of $h$ inside $\Gamma$ and $\mathcal{P}$ is the number of poles of $h$ inside $\Gamma$, including multiplicities.

Now, let $\Gamma$ be a piecewise differentiable closed curve in $\CC$ with positive orientation that does not pass through the point $z_0$. Then the \emph{index} of $z_0$ with respect to $\Gamma$, denoted $\text{Ind}_{\Gamma} (z_0)$, is defined by \cite[Page 115]{Ahlfors1979}
\begin{equation}\label{eq:Index}
\text{Ind}_{\Gamma} (z_0) = \frac{1}{2\pi i} \oint_{\Gamma} \frac{\mathrm{d}z}{z - z_0}.
\end{equation}
Note that $\text{Ind}_{\Gamma} (z_0)$ is also referred to as the \emph{winding number} of $\Gamma$ with respect to $z_0$. By substituting $z = h(\lambda)$ into Equation~\eqref{eq:Cauchy} and using Equation~\eqref{eq:Index} we arrive at the standard observation
\begin{equation}\label{eq:Cauchy2}
\mathcal{Z} - \mathcal{P} = \frac{1}{2\pi i} \oint_{h(\Gamma)} \frac{\mathrm{d}z}{z} = \text{Ind}_{h(\Gamma)} (0),
\end{equation}
where the term on the right of the last equality is the winding number of the closed curve $h(\Gamma)$ with respect to the origin. Therefore, the number of zeros minus the number of poles of $h$ inside $\Gamma$, including multiplicities, can be determined by calculating the winding number of the image $h(\Gamma)$ with respect to the origin.

Note that $G(\lambda)$ in Equation~\eqref{eq:Transcendental} is a holomorphic function, hence meromorphic, on $\CC$, and has no poles. Therefore, the number of zeros of $G(\lambda)$ inside a contour $\Gamma$ which satisifes the conditions for Equations~\eqref{eq:Cauchy} and \eqref{eq:Index} is equal to the winding number of $G(\Gamma)$ with respect to the origin. We will be considering rectangular contours with positive orientation in the right half-plane, and we need to know that $G(\lambda)$ is not identically zero in the region bounded by the closed contour. Our contour will bound an interval of the positive real axis arbitrarily close to, but excluding, the origin, and with arbitrary upper bound. For a contour which bounds sufficiently large positive real numbers, and by considering $\text{Re}(G(\lambda))$ on the positive real axis, we can see that $G(\lambda)$ is not identically zero in the region bounded by the contour. Furthermore, by \cite[Page 208, Theorem 10.18]{Rudin1986} it follows that the zeros of $G(\lambda)$ are isolated and countable, so we can always choose an appropriate rectangular contour which does not pass through a zero of $G(\lambda)$. In the Electronic Supplementary Material we graphically illustrate our application of Cauchy's argument principle. We now state and prove the required proposition.

\begin{proposition}\label{prop:Transcendental}
The transcendental characteristic equation $G(\lambda)$ in Equation~\eqref{eq:Transcendental} has at least one zero in $\CC$ with positive real part.
\end{proposition}

\begin{proof}
We consider three cases according to whether $2\alpha_S + 2\alpha_F - 3$ is negative, zero or positive.

Case 1: If $2\alpha_S + 2\alpha_F - 3 < 0$ then, since $G(0) = (2\alpha_S + 2\alpha_F - 3)r < 0$ and $\lim_{\lambda \to \infty} G \RestrictTo{\RR} (\lambda) = \lim_{\lambda \to \infty} \lambda^2 = \infty$, it follows that $G$ has a positive real zero by the intermediate value theorem.

Case 2: If $2\alpha_S + 2\alpha_F - 3 = 0$ then $G$ factors as $G(\lambda) = \lambda H(\lambda)$, where
\begin{align}
&H(\lambda) = \nonumber \\[0.0cm]
&\lambda - \bigg((2\alpha_S - 1) r \int_{0}^{U_S} e^{-\lambda r_F z} \, g_S (z) \, \mathrm{d}z + (2\alpha_F - 1) \int_{0}^{U_F} e^{-\lambda r_F z} \, g_F (z)\, \mathrm{d}z\bigg). \label{eq:FS15}
\end{align}
Since $2\alpha_S + 2\alpha_F - 3 = 0$ it follows that $2\alpha_S - 1 < 0$ implies $\alpha_F > 1$, so $2\alpha_S - 1 \ge 0$. Similarly, $2\alpha_F - 1 \ge 0$. Further, if both $2\alpha_S - 1 = 0$ and $2\alpha_F - 1 = 0$ then $2\alpha_S + 2\alpha_F - 2 = 0$, contradicting $2\alpha_S + 2\alpha_F - 3 = 0$. It therefore follows that $H(0) = -\big((2\alpha_S - 1)r + (2\alpha_F - 1)\big) < 0$. Since $\lim_{\lambda \to \infty} H \RestrictTo{\RR} (\lambda) = \lim_{\lambda \to \infty} \lambda = \infty$, it follows that $H$ has a positive real zero by the intermediate value theorem.

Case 3: Suppose now that $2\alpha_S + 2\alpha_F - 3 > 0$. We employ Cauchy's argument principle to show that the holomorphic function $G(\lambda)$ has a zero with positive real part. Let $\Gamma$ be the simple closed contour with positive orientation in the right half-plane of the complex plane defined piecewise as follows:
\begin{align}
\Gamma_1 &: \big(m (1-t) + (N/2) t\big) - i N, \quad 0 \le t \le 1, \label{eq:FS16} \\[0.1cm]
\Gamma_2 &: \big((N/2)(1-t) + (3N/2) t\big) - i N, \quad 0 \le t \le 1, \label{eq:FS17} \\[0.1cm]
\Gamma_3 &: (3N/2) + i \big((-N) (1-t) + Nt\big), \quad 0 \le t \le 1, \label{eq:FS18} \\[0.1cm]
\Gamma_4 &: \big((3N/2)(1-t) + (N/2) t\big) + i N, \quad 0 \le t \le 1, \label{eq:FS19} \\[0.1cm]
\Gamma_5 &: \big((N/2)(1-t) + m t\big) + i N, \quad 0 \le t \le 1, \label{eq:FS20} \\[0.1cm]
\Gamma_6 &: m + i \big(N (1-t) + (-N)t\big), \quad 0 \le t \le 1, \label{eq:FS21}
\end{align}
where we fix $m > 0$ to be arbitrarily small and $N > 0$  to be arbitrarily large. Note that $\Gamma = \bigcup_{j=1}^{6} \Gamma_j$ is rectangular, with vertices at $m-iN$, $3N/2-iN$, $3N/2+iN$, and $m+iN$. Since the zeros of a holomorphic function that is not identically zero are isolated and countable, we can choose $m$ arbitrarily small and $N$ arbitrarily large while ensuring $G(\lambda)$ is nonzero on $\Gamma$. Therefore, since $G(\lambda)$ has no poles, the number of zeros of $G(\lambda)$ inside $\Gamma$ is equal to the index of the image contour $G(\Gamma)$ with respect to the origin, $\text{Ind}_{G(\Gamma)} (0)$. We now begin our application of the argument principle. Since we only need to show the existence of one zero with a positive real part, it suffices to prove that $\text{Ind}_{G(\Gamma)} (0) \ge 1$. Specifically, we show that the image contour $G(\Gamma)$ crosses the positive real axis at least once in a counter-clockwise direction while encircling the origin, and doesn't cross the positive real axis in a clockwise direction.

We will traverse $\Gamma$ for one cycle in a counter-clockwise direction beginning with $\Gamma_1$, and determine when $G(\Gamma)$ crosses the positive real axis. For this it is helpful to consider the real and imaginary parts of $G(\lambda)$, so evaluating $G(\lambda)$ at the arbitrary complex number $\lambda = a + i b$ we have
\begin{align}
&\text{Re}(G(\lambda)) = a^2 - b^2 - a (2\alpha_S - 1) r \int_{0}^{U_S} e^{-a r_F z} \cos(b r_F z) \, g_S (z) \, \mathrm{d}z \nonumber \\[0.0cm]
&- a (2\alpha_F - 1) \int_{0}^{U_F} e^{- a r_F z} \cos(b r_F z) \, g_F (z) \, \mathrm{d}z \nonumber \\[0.0cm]
&- b (2\alpha_S - 1) r \int_{0}^{U_S} e^{-a r_F z} \sin(b r_F z) \, g_S (z) \, \mathrm{d}z \nonumber \\[0.0cm]
&- b (2\alpha_F - 1) \int_{0}^{U_F} e^{- a r_F z} \sin(b r_F z) \, g_F (z) \, \mathrm{d}z \nonumber \\[0.0cm]
&+ (2\alpha_S + 2\alpha_F - 3) r \int_{0}^{U_S} \int_{0}^{U_F} e^{- a r_F (z+v)} \cos(b r_F (z+v)) \, g_F (v) \, g_S (z) \, \mathrm{d}v \, \mathrm{d}z \, , \label {eq:FS22}
\end{align}
and
\begin{align}
&\text{Im}(G(\lambda)) = 2ab + a (2\alpha_S - 1) r \int_{0}^{U_S} e^{-a r_F z} \sin(b r_F z) \, g_S (z) \, \mathrm{d}z \nonumber \\[0.0cm]
&+ a (2\alpha_F - 1) \int_{0}^{U_F} e^{- a r_F z} \sin(b r_F z) \, g_F (z) \, \mathrm{d}z \nonumber \\[0.0cm]
&- b (2\alpha_S - 1) r \int_{0}^{U_S} e^{-a r_F z} \cos(b r_F z) \, g_S (z) \, \mathrm{d}z \nonumber \\[0.0cm]
&- b (2\alpha_F - 1) \int_{0}^{U_F} e^{- a r_F z} \cos(b r_F z) \, g_F (z) \, \mathrm{d}z \nonumber \\[0.0cm]
&- (2\alpha_S + 2\alpha_F - 3) r \int_{0}^{U_S} \int_{0}^{U_F} e^{- a r_F (z+v)} \sin(b r_F (z+v)) \, g_F (v) \, g_S (z) \, \mathrm{d}v \, \mathrm{d}z \, . \label{eq:FS23}
\end{align}
\noindent
In the following argument we shall generally use that $m$ is arbitrarily small and $N$ is arbitrarily large without further comment.

Consider $G(\lambda)$ along $\Gamma_1$, where $b = -N$ and $a$ increases from $m$ to $N/2$. For sufficiently large $N$ and for all $a \in [m,N/2]$, $\text{Re}(G(\lambda)) < 0$ and $\text{Re}(G(\lambda))$ is dominated by $N^2$. At the end of $\Gamma_1$ and for sufficiently large $N$, $\text{Im}(G(\lambda)) < 0$ and $\text{Im}(G(\lambda))$ is dominated by $N^2$. So $G(\Gamma_1)$ starts in the left half-plane and ends in the third quadrant.

Consider $G(\lambda)$ along $\Gamma_2$, where $b = -N$ and $a$ increases from $N/2$ to $3N/2$. At the end of $\Gamma_2$ and for sufficiently large $N$, $\text{Re}(G(\lambda)) > 0$ and $\text{Re}(G(\lambda))$ is dominated by $N^2$. For sufficiently large $N$ and for all $a \in [N/2,3N/2]$, $\text{Im}(G(\lambda)) < 0$ and $\text{Re}(G(\lambda))$ is dominated by $N^2$. So $G(\Gamma_2)$ starts in the third quadrant and ends in the fourth quadrant.

Consider $G(\lambda)$ along $\Gamma_3$, where $a = 3N/2$ and $b$ increases from $-N$ to $N$. For sufficiently large $N$, $\text{Re}(G(\lambda)) > 0$ and $\text{Re}(G(\lambda))$ is dominated by $N^2$. At the end of $\Gamma_3$ and for sufficiently large $N$, $\text{Im}(G(\lambda)) > 0$ and $\text{Im}(G(\lambda))$ is dominated by $N^2$. So $G(\Gamma_3)$ starts in the fourth quadrant and ends in the first quadrant. The image contour $G(\Gamma)$ has now crossed the positive real axis in a counter-clockwise direction.

Consider $G(\lambda)$ along $\Gamma_4$, where $b = N$ and $a$ decreases from $3N/2$ to $N/2$. At the end of $\Gamma_4$ and for sufficiently large $N$, $\text{Re}(G(\lambda)) < 0$ and $\text{Re}(G(\lambda))$ is dominated by $N^2$. For sufficiently large $N$ and for all $a \in [3N/2,N/2]$, $\text{Im}(G(\lambda)) > 0$ and $\text{Im}(G(\lambda))$ is dominated by $N^2$. So $G(\Gamma_4)$ starts in the first quadrant and ends in the second quadrant.

Consider $G(\lambda)$ along $\Gamma_5$, where $b = N$ and $a$ decreases from $N/2$ to $m$. At the end of $\Gamma_5$ and for sufficiently large $N$, $\text{Re}(G(\lambda)) < 0$ and $\text{Re}(G(\lambda))$ is dominated by $N^2$. $\text{Im}(G(\lambda))$ could be positive or negative. So $G(\Gamma_5)$ starts in the second quadrant and ends in the left half-plane.

Consider $G(\lambda)$ along $\Gamma_6$, where $a = m$ and $b$ decreases from $N$ to $-N$, which completes one circuit around $\Gamma$ in a counter-clockwise direction. If we fix $N$ to be as large as required then we can choose $m$ sufficiently small so that along $\Gamma_6$ the Equations~\eqref{eq:FS22} and \eqref{eq:FS23} are approximated arbitrarily closely by the equations
\begin{align}
\text{Re}(G(b)) = &- b^2 - b (2\alpha_S - 1) r \int_{0}^{U_S} \sin(b r_F z) \, g_S (z) \, \mathrm{d}z \nonumber \\[0.0cm]
&- b (2\alpha_F - 1) \int_{0}^{U_F} \sin(b r_F z) \, g_F (z) \, \mathrm{d}z \nonumber \\[0.0cm]
&+ (2\alpha_S + 2\alpha_F - 3) r \int_{0}^{U_S} \int_{0}^{U_F} \cos(b r_F (z+v)) \, g_F (v) \, g_S (z) \, \mathrm{d}v \, \mathrm{d}z \, , \label{eq:FS24}
\end{align}
\begin{align}
\text{Im}(G(b)) = &- b (2\alpha_S - 1) r \int_{0}^{U_S} \cos(b r_F z) \, g_S (z) \, \mathrm{d}z \nonumber \\[0.0cm]
&- b (2\alpha_F - 1) \int_{0}^{U_F} \cos(b r_F z) \, g_F (z) \, \mathrm{d}z \nonumber \\[0.0cm]
&- (2\alpha_S + 2\alpha_F - 3) r \int_{0}^{U_S} \int_{0}^{U_F} \sin(b r_F (z+v)) \, g_F (v) \, g_S (z) \, \mathrm{d}v \, \mathrm{d}z \, . \label{eq:FS25}
\end{align}
For notational simplicity, define the functions $f_1$, $f_2$, $f_3$, $g_1$, $g_2$, and $g_3$ by
\begingroup
\allowdisplaybreaks
\begin{align}
f_1 (b) &=  - b (2\alpha_S - 1) r \int_{0}^{U_S} \sin(b r_F z) \, g_S (z) \, \mathrm{d}z \, , \label{eq:FS26} \\[0.0cm]
f_2 (b) &=  - b (2\alpha_F - 1) \int_{0}^{U_F} \sin(b r_F z) \, g_F (z) \, \mathrm{d}z \, , \label{eq:FS27} \\[0.0cm]
f_3 (b) &=  (2\alpha_S + 2\alpha_F - 3) r \int_{0}^{U_S} \int_{0}^{U_F} \cos(b r_F (z+v)) \, g_F (v) \, g_S (z) \, \mathrm{d}v \, \mathrm{d}z \, , \label{eq:FS28} \\[0.0cm]
g_1 (b) &=  - b (2\alpha_S - 1) r \int_{0}^{U_S} \cos(b r_F z) \, g_S (z) \, \mathrm{d}z \, , \label{eq:FS29} \\[0.0cm]
g_2 (b) &=  - b (2\alpha_F - 1) \int_{0}^{U_F} \cos(b r_F z) \, g_F (z) \, \mathrm{d}z \, , \label{eq:FS30} \\[0.0cm]
g_3 (b) &=  - (2\alpha_S + 2\alpha_F - 3) r \int_{0}^{U_S} \int_{0}^{U_F} \sin(b r_F (z+v)) \, g_F (v) \, g_S (z) \, \mathrm{d}v \, \mathrm{d}z \, , \label{eq:FS31}
\end{align}
\endgroup
so that Equations~\eqref{eq:FS24} and \eqref{eq:FS25} become
\begin{align}
\text{Re}(G(b)) &= - b^2 + f_1 (b) + f_2 (b) + f_3 (b), \label{eq:FS32}\\[0.0cm]
\text{Im}(G(b)) &= g_1 (b) + g_2 (b) + g_3 (b). \label{eq:FS33}
\end{align}
Now, consider decreasing $b$ from $N$ to $-N$. The curves $f_1 (b) + i g_1 (b)$ and $f_2 (b) + i g_2 (b)$ have the same orientation as the spiral $-b \sin (b) - i b \cos(b)$, which is traversed counter-clockwise as $b$ decreases. Similarly, the curve $f_3 (b) + i g_3 (b)$ has the same orientation as the circle $\cos (b) - i \sin(b)$, which is also counter-clockwise. Note that for discrete delays the curves $f_1 (b) + i g_1 (b)$ and $f_2 (b) + i g_2 (b)$ are spirals and the curve $f_3 (b) + i g_3 (b)$ is a circle. The sum of these three curves, $\sum_{k=1}^{3} f_k (b) + i g_k (b)$, has counter-clockwise orientation, and the $-b^2$ term in $\text{Re}(G(b))$ translates these curves along the negative real axis. It follows that if $G(b)$ encircles the origin as $b$ decreases from $N$ to $-N$ then it does so in a counter-clockwise direction. In particular, $G(b)$ does not encircle the origin in a clockwise direction. So, for sufficiently small $m$, $G(\Gamma_6)$ can only encircle the origin in a counter-clockwise direction.

To ensure that the image contour $G(\Gamma)$ encircles the origin at least once, note that $G(\Gamma_6)$ crosses the positive real axis in a counter-clockwise direction at approximately the point $(2\alpha_S + 2\alpha_F - 3) r > 0$ for sufficiently small $m$, so it follows that $G(\Gamma_6)$ must cross the negative real axis at a point closer to the start of $\Gamma_6$. Therefore, the symmetry of $\text{Im}(G(\lambda))$ with respect to the real axis implies that $G(\Gamma)$ completes a cycle around the origin before the end of $G(\Gamma_6)$. We conclude that $\text{Ind}_{G(\Gamma)} (0) \ge 1$, and our proof is complete.
\end{proof}

\begin{proposition}[Equilibrium point $(1,0)$ when $\beta_S - \beta_F \ne 0$]\label{prop:Equil1}\hfill \\
Consider the system of delay differential equations~\eqref{eq:FS1} and \eqref{eq:FS2} with parameters, delay kernels, and history functions which satisfy \eqref{eq:param1}--\eqref{eq:history2}. For all $\beta_S$, $\beta_F \in \RR$, $(1,0)$ is locally stable when $\beta_S - \beta_F < 0$ and locally unstable when $\beta_S - \beta_F > 0$.
\end{proposition}

\begin{proof}
For $(S^{\text{*}},F^{\text{*}}) = (1,0)$, Equations~\eqref{eq:FS9} and \eqref{eq:FS10} become
\begin{align}
\frac{\mathrm{d} x(t)}{\mathrm{d} t} &= (2\alpha_S - 1) r \big(- x(t) - y(t)\big) - \beta y(t) \, , \label{eq:FS34} \\[0.1cm]
\frac{\mathrm{d} y(t)}{\mathrm{d} t} &= 2(1-\alpha_S) r \big(- x(t) - y(t)\big) + \beta y(t) \, . \label{eq:FS35}
\end{align}
Equations~\eqref{eq:FS34} and \eqref{eq:FS35} have a solution of the form in Equation~\eqref{eq:solution}, so substitution gives
\begin{equation}
e^{\lambda t} \begin{bmatrix} \lambda & 0 \\ 0 & \lambda \end{bmatrix} \begin{bmatrix} c_1\\ c_2 \end{bmatrix}
= \begin{bmatrix}
-(2\alpha_S - 1) r & \quad -(2\alpha_S - 1) r - \beta \\[0.1cm]
-2(1-\alpha_S) r & -2(1-\alpha_S) r + \beta
\end{bmatrix}
\begin{bmatrix} c_1\\ c_2 \end{bmatrix} e^{\lambda t}. \label{eq:FS36}
\end{equation}
To ensure that $(c_1 , c_2 )^{\intercal} \neq 0$ we must have the characteristic equation
\begin{equation}
G(\lambda) =
\begin{vmatrix}
-(2\alpha_S - 1) r - \lambda & -(2\alpha_S - 1) r - \beta \\[0.1cm]
-2(1-\alpha_S) r & -2(1-\alpha_S) r + \beta - \lambda
\end{vmatrix}
= \lambda^2 + \lambda (r-\beta) - \beta r \label{eq:FS37}
\end{equation}
equal to zero. The zeros of $G(\lambda)$ are the eigenvalues, given by
\begin{equation}
\lambda = -\frac{1}{2} (r-\beta) \pm \frac{1}{2} |\, r+\beta \,| = \text{$-r$, $\beta$}, \label{eq:FS38}
\end{equation}
and the result follows.
\end{proof}

\begin{proposition}[Equilibrium point $(0,1)$ when $\beta_S - \beta_F \ne 0$]\label{prop:Equil2}\hfill \\
Consider the system of delay differential equations~\eqref{eq:FS1} and \eqref{eq:FS2} with parameters, delay kernels, and history functions which satisfy \eqref{eq:param1}--\eqref{eq:history2}. For all $\beta_S$, $\beta_F \in \RR$, $(0,1)$ is locally stable when $\beta_S - \beta_F > 0$ and locally unstable when $\beta_S - \beta_F < 0$.
\end{proposition}

\begin{proof}
For $(S^{\text{*}},F^{\text{*}}) = (0,1)$, Equations~\eqref{eq:FS9} and \eqref{eq:FS10} become
\begin{align}
\frac{\mathrm{d} x(t)}{\mathrm{d} t} &= 2(1-\alpha_F) \big(- x(t) - y(t)\big) - \beta x(t) \, , \label{eq:FS39} \\[0.1cm]
\frac{\mathrm{d} y(t)}{\mathrm{d} t} &= (2\alpha_F - 1) \big(- x(t) - y(t)\big) + \beta x(t) \, . \label{eq:FS40}
\end{align}
Equations~\eqref{eq:FS39} and \eqref{eq:FS40} have a solution of the form in Equation~\eqref{eq:solution}, so substitution gives
\begin{equation}
e^{\lambda t} \begin{bmatrix} \lambda & 0 \\ 0 & \lambda \end{bmatrix} \begin{bmatrix} c_1\\ c_2 \end{bmatrix}
= \begin{bmatrix}
-2(1-\alpha_F) - \beta & \quad -2(1-\alpha_F) \\[0.1cm]
-(2\alpha_F - 1) + \beta & -(2\alpha_F - 1)
\end{bmatrix}
\begin{bmatrix} c_1\\ c_2 \end{bmatrix} e^{\lambda t}. \label{eq:FS41}
\end{equation}
To ensure that $(c_1 , c_2 )^{\intercal} \neq 0$ we must have the characteristic equation
\begin{equation}
G(\lambda)
= \begin{vmatrix}
-2(1-\alpha_F) - \beta - \lambda & -2(1-\alpha_F) \\[0.1cm]
-(2\alpha_F - 1) + \beta & -(2\alpha_F - 1) - \lambda
\end{vmatrix}
= \lambda^2 + \lambda (1+\beta) + \beta \label{eq:FS42}
\end{equation}
equal to zero. The zeros of $G(\lambda)$ are the eigenvalues, given by
\begin{equation}
\lambda = -\frac{1}{2} (1+\beta) \pm \frac{1}{2} |\, 1-\beta \,| = \text{$-1$, $-\beta$}, \label{eq:FS43}
\end{equation}
and the result follows.
\end{proof}

\begin{proposition}[Equilibrium points $(u,1-u)$ for $u \in {[0,1]}$ when $\beta_S = \beta_F$]\label{prop:Equil3}\hfill \\
Consider the system of delay differential equations~\eqref{eq:FS1} and \eqref{eq:FS2} with parameters, delay kernels, and history functions which satisfy \eqref{eq:param1}--\eqref{eq:history2}. For all $\beta_S$, $\beta_F \in \RR$ such that $\beta_S = \beta_F$ and for all $u \in [0,1]$ the equilibrium point $(u,1-u)$ is locally stable.
\end{proposition}

\begin{proof}
For $(S^{\text{*}},F^{\text{*}}) = (u,1 - u)$ with $u \in \RR$ and $\beta = 0$, Equations~\eqref{eq:FS9} and \eqref{eq:FS10} become
\begin{align}
\frac{\mathrm{d} x(t)}{\mathrm{d} t} &= \big((2\alpha_S - 1)ru + 2(1-\alpha_F)(1-u) \big) \big(- x(t) - y(t)\big) \, , \label{eq:FS44} \\[0.1cm]
\frac{\mathrm{d} y(t)}{\mathrm{d} t} &= \big(2(1-\alpha_S)ru + (2\alpha_F - 1)(1-u) \big) \big(- x(t) - y(t)\big) \, . \label{eq:FS45}
\end{align}
Equations~\eqref{eq:FS44} and \eqref{eq:FS45} have a solution of the form in Equation~\eqref{eq:solution}, so substitution gives
\begin{align}
e^{\lambda t} \begin{bmatrix} \lambda & 0 \\ 0 & \lambda \end{bmatrix} \begin{bmatrix} c_1\\ c_2 \end{bmatrix}
= \begin{bmatrix*}[l]
-(2\alpha_S - 1)ru & \quad -(2\alpha_S - 1)ru \\[0.1cm]
\quad - 2(1-\alpha_F)(1-u) & \qquad - 2(1-\alpha_F)(1-u) \\[0.3cm]
-2(1-\alpha_S)ru & \quad -2(1-\alpha_S)ru \\[0.1cm]
\quad - (2\alpha_F - 1)(1-u) & \qquad - (2\alpha_F - 1)(1-u)
\end{bmatrix*}
\begin{bmatrix} c_1\\ c_2 \end{bmatrix} e^{\lambda t}. \label{eq:FS46}
\end{align}
To ensure that $(c_1 , c_2 )^{\intercal} \neq 0$ we must have the characteristic equation
\begin{equation}
G(\lambda)
= \begin{vmatrix*}[l]
-(2\alpha_S - 1)ru - \lambda & \quad -(2\alpha_S - 1)ru \\[0.1cm]
\quad - 2(1-\alpha_F)(1-u) & \qquad - 2(1-\alpha_F)(1-u) \\[0.3cm]
-2(1-\alpha_S)ru & \quad -2(1-\alpha_S)ru - \lambda \\[0.1cm]
\quad - (2\alpha_F - 1)(1-u) & \qquad - (2\alpha_F - 1)(1-u)
\end{vmatrix*} = \lambda^2 - \lambda \big(u(1-r)-1\big) \label{eq:FS47}
\end{equation}
equal to zero. The zeros of $G(\lambda)$ are the eigenvalues, given by
\begin{equation}
\text{$\lambda = 0$, $u(1-r)-1$.} \label{eq:FS48}
\end{equation}
Since $r \in (0,1]$, for $u \in [0,1]$ we have $u(1-r)-1 < 0$, therefore $(u,1-u)$ is locally stable.
\end{proof}

\section{Supporting numerical simulations}\label{sec:Sim}
\begin{figure}
\centering
\includegraphics[width=0.8\textwidth]{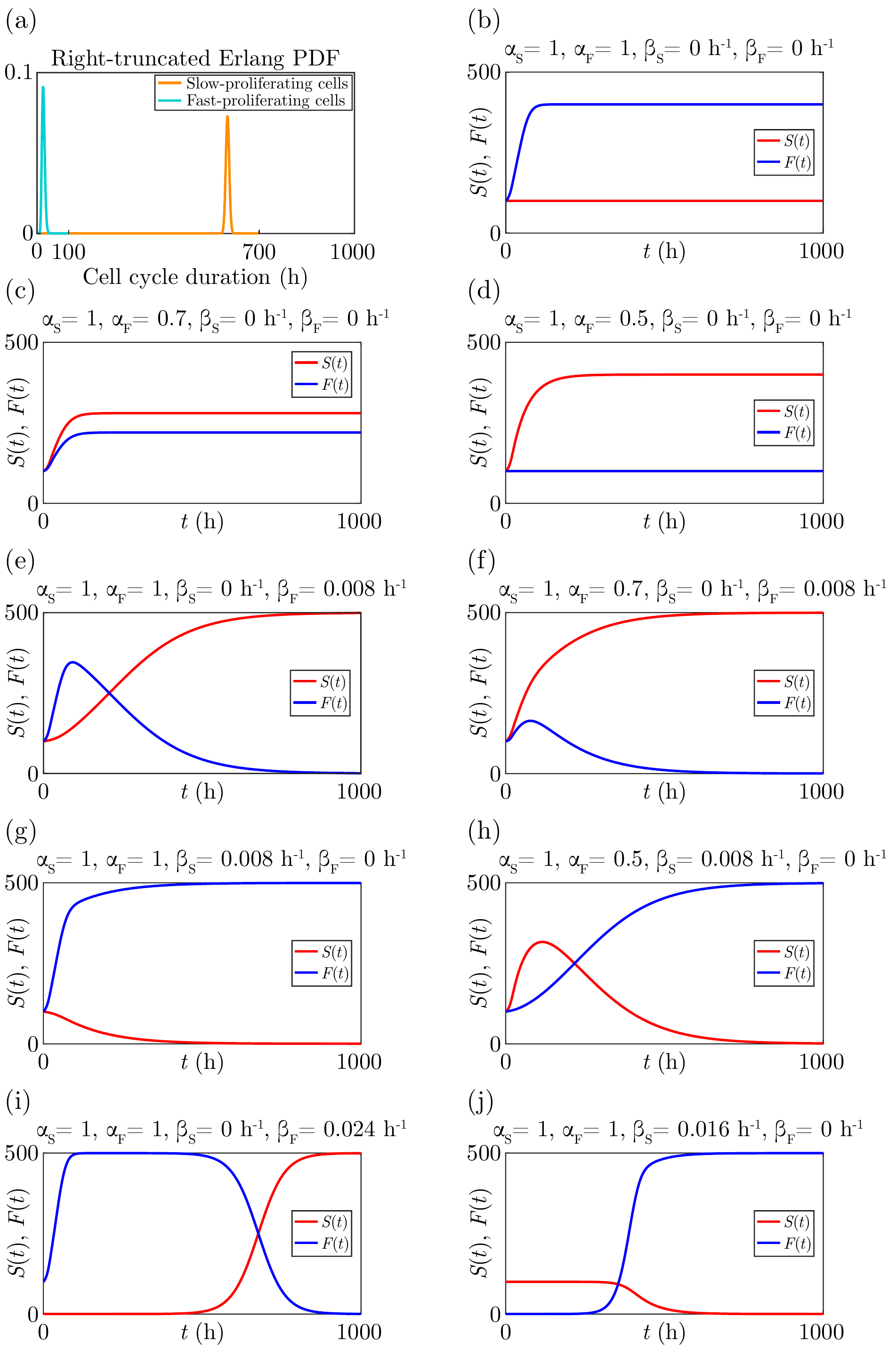}
\caption{Numerical simulations of our model in Equations~\eqref{eq:FS1} and \eqref{eq:FS2}. (a) Each delay kernel is the probability density function (PDF) of a right-truncated Erlang distribution (Electronic Supplementary Material). For slow-proliferating cells the Erlang density function has shape $k = 12000$ and rate $\lambda = 20$ h$^{-1}$ with mean 600 h, and is truncated at $U_S = 700$ h. For fast-proliferating cells the Erlang density function has shape $k = 20$ and rate $\lambda = 1$ h$^{-1}$ with mean 20 h, and is truncated at $U_F = 100$ h. (b)--(j) The simulations all use the parameters $K = 500$ and $r_S = r_F = 0.1$ h$^{-1}$. For (b)--(h) the history functions are $\phi_S (t) = 100 e^{r_S t}$ and $\phi_F (t) = 100 e^{r_F t}$, for (i) the history functions are $\phi_S (t) = 10^{-4} e^{r_S t}$ and $\phi_F (t) = 100 e^{r_F t}$, and for (j) the history functions are $\phi_S (t) = 100 e^{r_S t}$ and $\phi_F (t) = 10^{-4} e^{r_F t}$. Parameters specific to each simulation, namely $\alpha_S$, $\alpha_F$, $\beta_S$, and $\beta_F$, are indicated on the figure}
\label{fig:Fig3}
\end{figure}

We obtain numerical solutions of Equations~\eqref{eq:FS1} and \eqref{eq:FS2} using the forward Euler method, for which the temporal domain, $[0,1000]$ h, is uniformly discretised with a time step of duration $\Delta t = 0.1$ h. To approximate the distributed delays we use the trapezoidal rule with uniform discretisation of the integration interval into 500 subintervals. The distributed delays depend on past values of the functions $S(t)$ and $F(t)$, which are obtained by interpolating between the previously estimated values for $S(t)$ and $F(t)$. The interpolation is achieved using piecewise cubic Hermite interpolating polynomials, which are shape preserving. The sizes of the time step and the integration subintervals ensure grid-independence for our results. Examples of the simulations are shown in Figure~\ref{fig:Fig3}.

The delay kernels in our model are set as probability density functions of a right-truncated Erlang distribution (Electronic Supplementary Material), shown in Figure~\ref{fig:Fig3}(a). For slow-proliferating cells the Erlang density has shape $k = 12000$ and rate $\lambda = 20$ h$^{-1}$ with mean 600 h, and is truncated at $U_S = 700$ h. For fast-proliferating cells the Erlang density has shape $k = 20$ and rate $\lambda = 1$ h$^{-1}$ with mean 20 h, and is truncated at $U_F = 100$ h. All simulations use the parameters $K = 500$ and $r_S = r_F = 0.1$ h$^{-1}$. The parameters $\alpha_S$, $\alpha_F$, $\beta_S$, and $\beta_F$ are varied for the different simulations, as indicated in Figure~\ref{fig:Fig3}(b)--(h).

There are many options for the functional form of the history functions. One simple option is to use constant functions, however it is reasonable to assume that the cells grew exponentially in the past, so we use exponential functions with growth rates equal to the intrinsic growth rates of the slow- and fast-proliferating cells. The history functions are $\phi_S (t) = 100 e^{r_S \, t}$ for $t \in [-700,0]$ and $\phi_F (t) = 100 e^{r_F \, t}$ for $t \in [-100,0]$, for Figure~\ref{fig:Fig3}(b)--(h), $\phi_S (t) = 10^{-4} e^{r_S \, t}$ for $t \in [-700,0]$ and $\phi_F (t) = 100 e^{r_F \, t}$ for $t \in [-100,0]$, for Figure~\ref{fig:Fig3}(i), and $\phi_S (t) = 100 e^{r_S \, t}$ for $t \in [-700,0]$ and $\phi_F (t) = 10^{-4} e^{r_F \, t}$ for $t \in [-100,0]$, for Figure~\ref{fig:Fig3}(j). Since the state space is the function space $C$ in Equation~\eqref{eq:StateSpace}, choosing a different history function in $C$ results in a different solution. When $\beta_S \ne \beta_F$, different history functions may produce different transient dynamics, whereas the solutions have the same long-term behaviour. When $\beta_S = \beta_F$, however, there are infinitely many equilibrium points so different history functions can produce solutions with different long-term behaviour.

Figure~\ref{fig:Fig3}(b)--(d) shows simulations with $\beta_S = \beta_F = 0$ h$^{-1}$, so no induced switching between slow and fast proliferation. By Theorem~\ref{thrm:Stability} there are infinitely many locally-stable equilibrium points corresponding to the line segment between $(K,0)$ and $(0,K)$. The different equilibrium states are obtained by varying the levels of asymmetric division through $\alpha_S$ and $\alpha_F$, or using different history functions.

In Figure~\ref{fig:Fig3}(e)--(f) we show simulations with $\beta_S = 0$ h$^{-1}$ and $\beta_F = 0.008$ h$^{-1}$, therefore induced switching only from fast to slow proliferation. By Theorem~\ref{thrm:Stability} the equilibrium point $(K,0)$ is locally stable and the equilibrium point $(0,K)$ is locally unstable. In Figure~\ref{fig:Fig3}(g)--(h) we show simulations with $\beta_S = 0.008$ h$^{-1}$ and $\beta_F = 0$ h$^{-1}$, so induced switching only from slow to fast proliferation. By Theorem~\ref{thrm:Stability} the equilibrium point $(0,K)$ is locally stable and the equilibrium point $(K,0)$ is locally unstable. These simulations illustrate that induced switching determines the long-term behaviour of the solutions, while asymmetric division only influences the transient dynamics.

In Figure~\ref{fig:Fig3}(i)--(j) we set one of the history functions $\phi_S$ or $\phi_F$ close to zero over its domain, illustrating how a very small subpopulation can become the main subpopulation through induced switching, possibly requiring a long time period. It is particularly interesting that, in Figure~\ref{fig:Fig3}(i), the density of the fast-proliferating cells is very close to carrying capacity and appears to be at equilibrium for a long time, however through induced switching the slow-proliferating cells eventually become the main subpopulation.

\section{Discussion and outlook}\label{sec:Disc}
Proliferative heterogeneity in cancer cell populations constitutes a crucial challenge for cancer therapy, as slow-proliferating cells tend to be highly aggressive, have increased resistance to cytotoxic drugs, and can replenish the fast-proliferating subpopulation \cite{Ahn2017,Ahmed2018,Perego2018,Vallette2019}. The dynamics underlying tumour heterogeneity are not well understood, so improving cancer therapy depends on furthering this understanding \cite{Haass2014,Haass2015,Ahmed2018}. Theoretical approaches are well-placed to assist in elucidating the transient dynamics of intratumoural heterogeneity.

In this article we present a delay differential equation model for heterogeneous cell proliferation in which the total population consists of a subpopulation of slow-proliferating cells and a subpopulation of fast-proliferating cells. Our model incorporates the two cellular processes of asymmetric cell division and induced switching between proliferative states, which are important contributors to the dynamic heterogeneity of a cancer cell population \cite{Nelson2002,Bajaj2015,Dey-Guha2015,West2019}. The model is designed for investigating the transient dynamics of intratumoural heterogeneity with respect to cell proliferation. We employ delay differential equations in our model rather than ordinary differential equations in order to obtain transient dynamics consistent with the dynamics in a tumour. While the equilibrium states for our model are the same as those for the corresponding ordinary differential equations, the transient dynamics are very different, and model parameterisation with biologically-realistic values requires a model that incorporates realistic cell cycle durations for the slow- and fast-proliferating subpopulations.

Because the transient dynamics of a tumour are of primary interest, and the local stability analysis of our model provides only long-term behaviour near to the equilibrium, we must numerically simulate our model to explore the transient dynamics. We provide some examples of numerical simulations in Figure~\ref{fig:Fig3}, where we specify the delay kernels to be right-truncated Erlang distributions (Section 2 of Electronic Supplementary Material), and vary the parameters to demonstrate some of the possible dynamics within a tumour cell population. To exemplify some of the experimental scenarios to which our model is applicable, we consider a tumour that is treated with a cytotoxic drug which may induce cellular stress, causing the fast-proliferating cells to switch to the drug-resistant slow-proliferating phenotype \cite{Ahmed2018} through the mechanisms of asymmetric cell division and induced switching.

We show simulations where there is no induced switching between the slow- and fast-proliferating subpopulations in Figure~\ref{fig:Fig3}(b)--(d). If tumour cells experience no microenvironmental stress, then all cell divisions may be symmetric, corresponding to the situation in Figure~\ref{fig:Fig3}(b) where the fast-proliferating cells rapidly populate the tumour until the total cell density reaches carrying capacity. If a drug is introduced into the tumour microenvironment then the fast-proliferating cells may experience cellular stress, inducing the fast-proliferating cells into asymmetric cell division as a survival strategy, as the slow-proliferating phenotype is drug resistant. Figure~\ref{fig:Fig3}(c)--(d) illustrates this behaviour, first for an intermediate level of asymmetric division in (c), and then for the maximum level of asymmetric division in (d).

Figure~\ref{fig:Fig3}(e) shows a simulation where fast-proliferating cells are under stress due to the presence of a drug, and are induced to switch to slow proliferation through signals from slow-proliferating cells as a survival strategy. Alternatively, the induced signalling may arise from the highly invasive slow-proliferating cells \cite{Chapman2014,Ahmed2018} influencing the less invasive fast-proliferating cells to switch to the more invasive slow-proliferating phenotype. In our simulation the total cell population appears to reach the carrying capacity density at around 200 hours, however the dynamics of induced switching of cells from fast to slow proliferation continues until all cells are slow proliferating. This is important because, while tumour growth has effectively ceased, the tumour is becoming increasingly drug resistant and invasive over time until the whole tumour is composed of drug resistant and invasive cells. Therefore, effective early treatment of the tumour is required in order to prevent the tumour from becoming more aggressive and treatment resistant. If the per capita interaction strength of slow-proliferating cells to induce fast-proliferating cells to switch to slow proliferation is obtained experimentally, then our model could be used to predict the transient change in the proportion of slow-proliferating cells in the population, and therefore the changes in invasiveness and drug resistance of the tumour.

Now consider a tumour composed of mostly fast-proliferating cells and a very small proportion of slow-proliferating cells, as in Figure~\ref{fig:Fig3}(i). The fast-proliferating cells undergo induced switching to slow proliferation, perhaps due to stress from an introduced drug or to increase invasiveness. Initially the fast-proliferating subpopulation grows to reach near the carry capacity density, and the system appears to be in equilibrium for an extended period of time. As the tumour is almost completely composed of fast-proliferating cells, it is in the least invasive and most drug sensitive state. Without knowledge of the presence of induced switching, experimental investigations may not reveal that the tumour is in the process of becoming highly invasive and drug resistant. Indeed, once the density of the slow-proliferating cells has reached a sufficient but still very low level, the tumour rapidly becomes populated by slow-proliferating cells through induced switching of the fast-proliferating cells.

Finally, consider a small tumour comprised mostly of slow-proliferating cells and a very small proportion of fast-proliferating cells, as in Figure~\ref{fig:Fig3}(j). Signals from the fast-proliferating cells induce the slow-proliferating cells to switch to fast proliferation. Experimentally, this could correspond to a tumour that has been treated with a drug which caused the death of most of the fast-proliferating cells. For an extended period of time the tumour grows very little, until the density of fast-proliferating cells is high enough that the induced switching from slow- to fast-proliferation rapidly grows the tumour to the maximum sustainable size. Our model is therefore able to provide an estimate of tumour growth over time following drug treatment, when the cells can undergo induced switching.

There are numerous possibilities for future work. Induced switching between proliferative states could take many forms. In tumours the slow-proliferating state may continually arise and disappear \cite{Roesch2010}, so it would be interesting to accommodate time-dependent induced switching into the model, which could be either periodic or aperiodic. We could also consider the induced switching to have an explicit dependence on density, so that no switching occurs from a particular proliferative state when the density of cells from the other proliferative state is above a certain value. A similar explicit density dependence could be implemented for asymmetric cell division, which occurs at constant proportions in our current model. These explicit dependences on density may be relevant for slow-proliferating subpopulations in tumours that appear to persist over time and maintain the relative size of the subpopulation \cite{Perego2018,Vallette2019}. Our model could also be extended to include the additional process of spontaneous switching between proliferative states, which is independent of other cells and may be stochastic.

While our model has implicit spatial structure, since the dependent variables are cell densities, we could include spatial structure explicitly. We could then explicitly model cell migration with a diffusive term \cite{Vittadello2018}. Further, induced switching could be modelled as a more localised process where the rate of a cell switching proliferative states is determined by the density of cells in a different proliferative state within a given radius of the cell, where the interaction strength decreases with distance from the cell.

We could also extend our model to more than two dependent variables. For example, we could consider fast-, slow-, and very-slow-proliferating subpopulations. Another possible extension is to include apoptosis. Much of our analysis in this article is likely to be easily generalised to an extended version of our model. The more challenging aspect could be the analysis of the corresponding transcendental characteristic equations, however taking a more abstract approach for an extended model with an arbitrary $n$ dependent variables could simplify the problem.

\vspace{1cm}

\section*{Code availability}
The code for the algorithm to replicate the numerical simulations in this work is available on GitHub at https://github.com/DrSeanTVittadello/Vittadello2020.

\section*{Acknowledgements}
NKH is a Cameron fellow of the Melanoma and Skin Cancer Research Institute, and is supported by the National Health and Medical Research Council of Australia (APP1084893). MJS is supported by the Australian Research Council (DP170100474).

\section*{Conflict of interest}
The authors declare that they have no conflict of interest.

\bibliographystyle{apalike}
\bibliography{SeanBibliography}

\newcommand{\noop}[1]{}
\begin{thebibliography}{}

\bibitem[Ahlfors, 1979]{Ahlfors1979}
Ahlfors, L.~V. (1979).
\newblock {\em Complex Analysis}.
\newblock McGraw-Hill, third edition.

\bibitem[Ahmed and Haass, 2018]{Ahmed2018}
Ahmed, F. and Haass, N.~K. (2018).
\newblock Microenvironment-driven dynamic heterogeneity and phenotypic
  plasticity as a mechanism of melanoma therapy resistance.
\newblock {\em Frontiers in Oncology}, 8:173.

\bibitem[Ahn et~al., 2017]{Ahn2017}
Ahn, A., Chatterjee, A., and Eccles, M.~R. (2017).
\newblock The slow cycling phenotype: A growing problem for treatment
  resistance in melanoma.
\newblock {\em Molecular Cancer Therapeutics}, 16:1002--1009.

\bibitem[Arino, 1995]{Arino1995}
Arino, O. (1995).
\newblock A survey of structured cell population dynamics.
\newblock {\em Acta Biotheoretica}, 43:3--25.

\bibitem[Arino and Kimmel, 1989]{Arino1989}
Arino, O. and Kimmel, M. (1989).
\newblock Asymptotic behavior of a nonlinear functional-integral equation of
  cell kinetics with unequal division.
\newblock {\em Journal of Mathematical Biology}, 27:341--354.

\bibitem[Bajaj et~al., 2015]{Bajaj2015}
Bajaj, J., Zimdahl, B., and Reya, T. (2015).
\newblock Fearful symmetry: subversion of asymmetric division in cancer
  development and progression.
\newblock {\em Cancer Research}, 75:792--797.

\bibitem[Baker et~al., 1997]{Baker1997}
Baker, C. T.~H., Bocharov, G.~A., and Paul, C. A.~H. (1997).
\newblock Mathematical modelling of the interleukin-2 {T}-cell system: a
  comparative study of approaches based on ordinary and delay differential
  equation.
\newblock {\em Journal of Theoretical Medicine}, 1:117--128.

\bibitem[Baker et~al., 1998]{Baker1998}
Baker, C. T.~H., Bocharov, G.~A., Paul, C. A.~H., and Rihan, F.~A. (1998).
\newblock Modelling and analysis of time-lags in some basic patterns of cell
  proliferation.
\newblock {\em Journal of Mathematical Biology}, 37:341--371.

\bibitem[Beaumont et~al., 2016]{Beaumont2016}
Beaumont, K.~A., Hill, D.~S., Daignault, S.~M., Lui, G.~Y., Sharp, D.~M.,
  Gabrielli, B., Weninger, W., and Haass, N.~K. (2016).
\newblock Cell cycle phase-specific drug resistance as an escape mechanism of
  melanoma cells.
\newblock {\em Journal of Investigative Dermatology}, 136:1479--1489.

\bibitem[Billy et~al., 2014]{Billy2014}
Billy, F., Clairambaultt, J., Fercoq, O., Gaubertt, S., Lepoutre, T., Ouillon,
  T., and Saito, S. (2014).
\newblock Synchronisation and control of proliferation in cycling cell
  population models with age structure.
\newblock {\em Mathematics and Computers in Simulation}, 96:66--94.

\bibitem[Byrne and Drasdo, 2009]{Byrne2009}
Byrne, H. and Drasdo, D. (2009).
\newblock Individual-based and continuum models of growing cell populations: a
  comparison.
\newblock {\em Journal of Mathematical Biology}, 58:657--687.

\bibitem[Byrne, 1997]{Byrne1997}
Byrne, H.~M. (1997).
\newblock The effect of time delays on the dynamics of avascular tumor growth.
\newblock {\em Mathematical Biosciences}, 144:83--117.

\bibitem[Cai et~al., 2007]{Cai2007}
Cai, A.~Q., Landman, K.~A., and Hughes, B.~D. (2007).
\newblock Multi-scale modeling of a wound-healing cell migration assay.
\newblock {\em Journal of Theoretical Biology}, 245:576--594.

\bibitem[Cassidy et~al., 2019]{Cassidy2019}
Cassidy, T., Craig, M., and Humphries, A.~R. (2019).
\newblock Equivalences between age structured models and state dependent
  distributed delay differential equations.
\newblock {\em Mathematical Biosciences and Engineering}, 16:5419--5450.

\bibitem[Cassidy and Humphries, 2020]{Cassidy2020}
Cassidy, T. and Humphries, A.~R. (2020).
\newblock A mathematical model of viral oncology as an immuno-oncology
  instigator.
\newblock {\em Mathematical Medicine and Biology}, 37:117--151.

\bibitem[Chao et~al., 2019]{Chao2019}
Chao, H.~X., Fakhreddin, R.~I., Shimerov, H.~K., Kedziora, K.~M., Kumar, R.~J.,
  Perez, J., Limas, J.~C., Grant, G.~D., Cook, J.~G., Gupta, G.~P., and Purvis,
  J.~E. (2019).
\newblock Evidence that the human cell cycle is a series of uncoupled,
  memoryless phases.
\newblock {\em Molecular Systems Biology}, 15:e8604.

\bibitem[Chapman et~al., 2014]{Chapman2014}
Chapman, A., Fernandez~del Ama, L., Ferguson, J., Kamarashev, J., Wellbrock,
  C., and Hurlstone, A. (2014).
\newblock Heterogeneous tumor subpopulations cooperate to drive invasion.
\newblock {\em Cell Reports}, 8:688--695.

\bibitem[Clairambault and Fercoq, 2016]{Clairambault2016}
Clairambault, J. and Fercoq, O. (2016).
\newblock Physiologically structured cell population dynamic models with
  applications to combined drug delivery optimisation in oncology.
\newblock {\em Mathematical Modelling of Natural Phenomena}, 11:45--70.

\bibitem[Dey-Guha et~al., 2015]{Dey-Guha2015}
Dey-Guha, I., Alves, C.~P., Yeh, A.~C., Salony, Sole, X., Darp, R., and
  Ramaswamy, S. (2015).
\newblock A mechanism for asymmetric cell division resulting in proliferative
  asynchronicity.
\newblock {\em Molecular Cancer Research}, 13:223--230.

\bibitem[Dey-Guha et~al., 2011]{Dey-Guha2011}
Dey-Guha, I., Wolfer, A., Yeh, A.~C., Albeck, J.~G., Darp, R., Leon, E.,
  Wulfkuhle, J., Petricoin, E.~F., Wittner, B.~S., and Ramaswamy, S. (2011).
\newblock Asymmetric cancer cell division regulated by {AKT}.
\newblock {\em Proceedings of the National Academy of Sciences of the United
  States of America}, 108:12845--12850.

\bibitem[Diekmann et~al., 1995]{Diekmann1995}
Diekmann, O., van Gils, S.~A., Verduyn~Lunel, S.~M., and Walther, H.-O. (1995).
\newblock {\em Delay Equations}.
\newblock Springer New York.

\bibitem[Engelborghs et~al., 2000]{Engelborghs2000}
Engelborghs, K., Luzyanina, T., and Roose, D. (2000).
\newblock Numerical bifurcation analysis of delay differential equations.
\newblock {\em Journal of Computational and Applied Mathematics}, 125:265--275.

\bibitem[Gabriel et~al., 2012]{Gabriel2012}
Gabriel, P., Garbett, S.~P., Quaranta, V., Tyson, D.~R., and Webb, G.~F.
  (2012).
\newblock The contribution of age structure to cell population responses to
  targeted therapeutics.
\newblock {\em Journal of Theoretical Biology}, 311:19--27.

\bibitem[Gallaher et~al., 2019]{Gallaher2019}
Gallaher, J.~A., Brown, J.~S., and Anderson, A. R.~A. (2019).
\newblock The impact of proliferation-migration tradeoffs on phenotypic
  evolution in cancer.
\newblock {\em Scientific Reports}, 9:2425.

\bibitem[Gavagnin et~al., 2019]{Gavagnin2019}
Gavagnin, E., Ford, M.~J., Mort, R.~L., Rogers, T., and Yates, C.~A. (2019).
\newblock The invasion speed of cell migration models with realistic cell cycle
  time distributions.
\newblock {\em Journal of Theoretical Biology}, 481:91--99.

\bibitem[Getto et~al., 2019]{Getto2019}
Getto, P., Gyllenberg, M., Nakata, Y., and Scarabel, F. (2019).
\newblock Stability analysis of a state-dependent delay differential equation
  for cell maturation: analytical and numerical methods.
\newblock {\em Journal of Mathematical Biology}, 79:281--328.

\bibitem[Getto and Waurick, 2016]{Getto2016}
Getto, P. and Waurick, M. (2016).
\newblock A differential equation with state-dependent delay from cell
  population biology.
\newblock {\em Journal of Differential Equations}, 260:6176--6200.

\bibitem[Greene et~al., 2015]{Greene2015}
Greene, J.~M., Levy, D., Fung, K.~L., Souza, P.~S., Gottesman, M.~M., and Lavi,
  O. (2015).
\newblock Modeling intrinsic heterogeneity and growth of cancer cells.
\newblock {\em Journal of Theoretical Biology}, 367:262--277.

\bibitem[Haass, 2015]{Haass2015}
Haass, N.~K. (2015).
\newblock Dynamic tumor heterogeneity in melanoma therapy: how do we address
  this in a novel model system?
\newblock {\em Melanoma Management}, 2:93--95.

\bibitem[Haass et~al., 2014]{Haass2014}
Haass, N.~K., Beaumont, K.~A., Hill, D.~S., Anfosso, A., Mrass, P., Munoz,
  M.~A., Kinjyo, I., and Weninger, W. (2014).
\newblock Real-time cell cycle imaging during melanoma growth, invasion, and
  drug response.
\newblock {\em Pigment Cell {\&} Melanoma Research}, 27:764--776.

\bibitem[Hanahan and Weinberg, 2011]{Hanahan2011}
Hanahan, D. and Weinberg, R.~A. (2011).
\newblock Hallmarks of cancer: The next generation.
\newblock {\em Cell}, 144:646--674.

\bibitem[Huang et~al., 2016]{Huang2016}
Huang, C., Cao, J., Wen, F., and Yang, X. (2016).
\newblock Stability analysis of {SIR} model with distributed delay on complex
  networks.
\newblock {\em {PLOS} {ONE}}, 11:e0158813.

\bibitem[Jin et~al., 2018]{Jin2018}
Jin, W., McCue, S.~W., and Simpson, M.~J. (2018).
\newblock Extended logistic growth model for heterogeneous populations.
\newblock {\em Journal of Theoretical Biology}, 445:51--61.

\bibitem[Kaslik and Neamtu, 2018]{Kaslik2018}
Kaslik, E. and Neamtu, M. (2018).
\newblock Stability and hopf bifurcation analysis for the
  hypothalamic-pituitary-adrenal axis model with memory.
\newblock {\em Mathematical Medicine and Biology}, 35:49--78.

\bibitem[Khasawneh and Mann, 2011]{Khasawneh2011}
Khasawneh, F.~A. and Mann, B.~P. (2011).
\newblock Stability of delay integro-differential equations using a spectral
  element method.
\newblock {\em Mathematical and Computer Modelling}, 54:2493--2503.

\bibitem[Kuang, 1993]{Kuang1993}
Kuang, Y. (1993).
\newblock {\em Delay Differential Equations: With Applications in Population
  Dynamics}.
\newblock Academic Press.

\bibitem[Lebowitz and Rubinow, 1974]{Lebowitz1974}
Lebowitz, J.~L. and Rubinow, S.~I. (1974).
\newblock A theory for the age and generation time distribution of a microbial
  population.
\newblock {\em Journal of Mathematical Biology}, 1:17--36.

\bibitem[Lu, 1991]{Lu1991}
Lu, L. (1991).
\newblock Numerical stability of the $\theta$-methods for systems of
  differential equations with several delay terms.
\newblock {\em Journal of Computational and Applied Mathematics}, 34:291--304.

\bibitem[Mackey and Rudnicki, 1994]{Mackey1994}
Mackey, M.~C. and Rudnicki, R. (1994).
\newblock Global stability in a delayed partial differential equation
  describing cellular replication.
\newblock {\em Journal of Mathematical Biology}, 33:89--109.

\bibitem[Maini et~al., 2004]{Maini2004a}
Maini, P.~K., McElwain, D. L.~S., and Leavesley, D.~I. (2004).
\newblock Traveling wave model to interpret a wound-healing cell migration
  assay for human peritoneal mesothelial cells.
\newblock {\em Tissue Engineering}, 10:475--482.

\bibitem[Matson and Cook, 2017]{Matson2017}
Matson, J.~P. and Cook, J.~G. (2017).
\newblock Cell cycle proliferation decisions: the impact of single cell
  analyses.
\newblock {\em The {FEBS} Journal}, 284:362--375.

\bibitem[McClatchey and Yap, 2012]{McClatchey2012}
McClatchey, A.~I. and Yap, A.~S. (2012).
\newblock Contact inhibition (of proliferation) redux.
\newblock {\em Current Opinion in Cell Biology}, 24:685--694.

\bibitem[McCluskey, 2010]{McCluskey2010}
McCluskey, C.~C. (2010).
\newblock Global stability of an $sir$ epidemic model with delay and general
  nonlinear incidence.
\newblock {\em Mathematical Biosciences and Engineering}, 7:837--850.

\bibitem[Moore and Lyle, 2011]{Moore2011}
Moore, N. and Lyle, S. (2011).
\newblock Quiescent, slow-cycling stem cell populations in cancer: A review of
  the evidence and discussion of significance.
\newblock {\em Journal of Oncology}, 2011:396076.

\bibitem[Nelson and Chen, 2002]{Nelson2002}
Nelson, C.~M. and Chen, C.~S. (2002).
\newblock Cell--cell signaling by direct contact increases cell proliferation
  via a {PI3K}-dependent signal.
\newblock {\em {FEBS} Letters}, 514:238--242.

\bibitem[Pavel et~al., 2018]{Pavel2018}
Pavel, M., Renna, M., Park, S.~J., Menzies, F.~M., Ricketts, T., Füllgrabe,
  J., Ashkenazi, A., Frake, R.~A., Lombarte, A.~C., Bento, C.~F., Franze, K.,
  and Rubinsztein, D.~C. (2018).
\newblock Contact inhibition controls cell survival and proliferation via
  {YAP}/{TAZ}-autophagy axis.
\newblock {\em Nature Communications}, 9:2961.

\bibitem[Perego et~al., 2018]{Perego2018}
Perego, M., Maurer, M., Wang, J.~X., Shaffer, S., M{\" u}ller, A.~C.,
  Parapatics, K., Li, L., Hristova, D., Shin, S., Keeney, F., Liu, S., Xu, X.,
  Raj, A., Jensen, J.~K., Bennett, K.~L., Wagner, S.~N., Somasundaram, R., and
  Herlyn, M. (2018).
\newblock A slow-cycling subpopulation of melanoma cells with highly invasive
  properties.
\newblock {\em Oncogene}, 37:302--312.

\bibitem[Puliafito et~al., 2012]{Puliafito2012}
Puliafito, A., Hufnagel, L., Neveu, P., Streichan, S., Sigal, A., Fygenson,
  D.~K., and Shraiman, B.~I. (2012).
\newblock Collective and single cell behavior in epithelial contact inhibition.
\newblock {\em Proceedings of the National Academy of Sciences of the United
  States of America}, 109:739--744.

\bibitem[Roesch et~al., 2010]{Roesch2010}
Roesch, A., Fukunaga-Kalabis, M., Schmidt, E.~C., Zabierowski, S.~E., Brafford,
  P.~A., Vultur, A., Basu, D., Gimotty, P., Vogt, T., and Herlyn, M. (2010).
\newblock A temporarily distinct subpopulation of slow-cycling melanoma cells
  is required for continuous tumor growth.
\newblock {\em Cell}, 141:583--594.

\bibitem[Rudin, 1986]{Rudin1986}
Rudin, W. (1986).
\newblock {\em Real and Complex Analysis}.
\newblock McGraw-Hill Education, third edition.

\bibitem[Sakaue-Sawano et~al., 2008]{Sakaue_Sawano2008}
Sakaue-Sawano, A., Kurokawa, H., Morimura, T., Hanyu, A., Hama, H., Osawa, H.,
  Kashiwagi, S., Fukami, K., Miyata, T., Miyoshi, H., Imamura, T., Ogawa, M.,
  Masai, H., and Miyawaki, A. (2008).
\newblock Visualizing spatiotemporal dynamics of multicellular cell-cycle
  progression.
\newblock {\em Cell}, 132:487--498.

\bibitem[Sarapata and de~Pillis, 2014]{Sarapata2014}
Sarapata, E.~A. and de~Pillis, L.~G. (2014).
\newblock A comparison and catalog of intrinsic tumor growth models.
\newblock {\em Bulletin of Mathematical Biology}, 76:2010--2024.

\bibitem[Scott et~al., 2013]{Scott2013}
Scott, J.~G., Basanta, D., Anderson, A. R.~A., and Gerlee, P. (2013).
\newblock A mathematical model of tumour self-seeding reveals secondary
  metastatic deposits as drivers of primary tumour growth.
\newblock {\em Journal of the Royal Society Interface}, 10:20130011.

\bibitem[Sherratt and Murray, 1990]{Sherratt1990}
Sherratt, J.~A. and Murray, J.~D. (1990).
\newblock Models of epidermal wound healing.
\newblock {\em Proceedings of the Royal Society B}, 241:29--36.

\bibitem[Simpson et~al., 2018]{Simpson2018}
Simpson, M.~J., Jin, W., Vittadello, S.~T., Tambyah, T.~A., Ryan, J.~M.,
  Gunasingh, G., Haass, N.~K., and McCue, S.~W. (2018).
\newblock Stochastic models of cell invasion with fluorescent cell cycle
  indicators.
\newblock {\em Physica A: Statistical Mechanics and its Applications},
  510:375--386.

\bibitem[Smalley and Herlyn, 2009]{Smalley2009}
Smalley, K. S.~M. and Herlyn, M. (2009).
\newblock Integrating tumor-initiating cells into the paradigm for melanoma
  targeted therapy.
\newblock {\em International Journal of Cancer}, 124:1245--1250.

\bibitem[Smith, 2011]{Smith2011}
Smith, H. (2011).
\newblock {\em An Introduction to Delay Differential Equations with
  Applications to the Life Sciences}.
\newblock Springer-Verlag GmbH.

\bibitem[Spoerri et~al., 2017]{Spoerri2017}
Spoerri, L., Beaumont, K.~A., Anfosso, A., and Haass, N.~K. (2017).
\newblock Real-time cell cycle imaging in a 3d cell culture model of melanoma.
\newblock {\em Methods in Molecular Biology}, 1612:401--416.

\bibitem[Sun, 2006]{Sun2006}
Sun, L. (2006).
\newblock Stability analysis for delay differential equations with multidelays
  and numerical examples.
\newblock {\em Mathematics of Computation}, 75:151--165.

\bibitem[Swanson et~al., 2003]{Swanson2003}
Swanson, K.~R., Bridge, C., Murray, J., and Alvord, E.~C. (2003).
\newblock Virtual and real brain tumors: using mathematical modeling to
  quantify glioma growth and invasion.
\newblock {\em Journal of the Neurological Sciences}, 216:1--10.

\bibitem[Vallette et~al., 2019]{Vallette2019}
Vallette, F.~M., Olivier, C., L{\'{e}}zot, F., Oliver, L., Cochonneau, D.,
  Lalier, L., Cartron, P.-F., and Heymann, D. (2019).
\newblock Dormant, quiescent, tolerant and persister cells: Four synonyms for
  the same target in cancer.
\newblock {\em Biochemical Pharmacology}, 162:169--176.

\bibitem[Vermeulen et~al., 2003]{Vermeulen2003}
Vermeulen, K., {Van Bockstaele}, D.~R., and Berneman, Z.~N. (2003).
\newblock The cell cycle: a review of regulation, deregulation and therapeutic
  targets in cancer.
\newblock {\em Cell Proliferation}, 36:131--149.

\bibitem[Villasana and Radunskaya, 2003]{Villasana2003}
Villasana, M. and Radunskaya, A. (2003).
\newblock A delay differential equation model for tumor growth.
\newblock {\em Journal of Mathematical Biology}, 47:270--294.

\bibitem[Vittadello et~al., 2018]{Vittadello2018}
Vittadello, S.~T., McCue, S.~W., Gunasingh, G., Haass, N.~K., and Simpson,
  M.~J. (2018).
\newblock Mathematical models for cell migration with real-time cell cycle
  dynamics.
\newblock {\em Biophysical Journal}, 114:1241--1253.

\bibitem[Vittadello et~al., 2019]{Vittadello2019}
Vittadello, S.~T., McCue, S.~W., Gunasingh, G., Haass, N.~K., and Simpson,
  M.~J. (2019).
\newblock Mathematical models incorporating a multi-stage cell cycle replicate
  normally-hidden inherent synchronization in cell proliferation.
\newblock {\em Journal of the Royal Society Interface}, 16:20190382.

\bibitem[Vittadello et~al., 2020]{Vittadello2020}
Vittadello, S.~T., McCue, S.~W., Gunasingh, G., Haass, N.~K., and Simpson,
  M.~J. (2020).
\newblock Examining go-or-grow using fluorescent cell-cycle indicators and
  cell-cycle-inhibiting drugs.
\newblock {\em Biophysical Journal}, 118:1243--1247.

\bibitem[Webb, 1986]{Webb1986}
Webb, G.~F. (1986).
\newblock A model of proliferating cell populations with inherited cycle
  length.
\newblock {\em Journal of Mathematical Biology}, 23:269--282.

\bibitem[Weber et~al., 2014]{Weber2014}
Weber, T.~S., Jaehnert, I., Schichor, C., Or-Guil, M., and Carneiro, J. (2014).
\newblock Quantifying the length and variance of the eukaryotic cell cycle
  phases by a stochastic model and dual nucleoside pulse labelling.
\newblock {\em {PLoS} Computational Biology}, 10:e1003616.

\bibitem[West and Newton, 2019]{West2019}
West, J. and Newton, P.~K. (2019).
\newblock Cellular interactions constrain tumor growth.
\newblock {\em Proceedings of the National Academy of Sciences of the United
  States of America}, 116:1918--1923.

\bibitem[Yates et~al., 2017]{Yates2017}
Yates, C.~A., Ford, M.~J., and Mort, R.~L. (2017).
\newblock A multi-stage representation of cell proliferation as a markov
  process.
\newblock {\em Bulletin of Mathematical Biology}, 79:2905--2928.

\bibitem[Zhu and Thompson, 2019]{Zhu2019}
Zhu, J. and Thompson, C.~B. (2019).
\newblock Metabolic regulation of cell growth and proliferation.
\newblock {\em Nature Reviews. Molecular Cell Biology}, 20:436--450.

\end{thebibliography}


 \newcommand{\noop}[1]{}
\begin{thebibliography}{10}

\bibitem{Vittadello2020}
Vittadello ST, McCue SW, Gunasingh G, Haass NK, Simpson MJ.
\newblock Examining go-or-grow using fluorescent cell-cycle indicators and
  cell-cycle-inhibiting drugs.
\newblock Biophysical Journal. 2020;118:1243--1247.

\bibitem{Haass2014}
Haass NK, Beaumont KA, Hill DS, Anfosso A, Mrass P, Munoz MA, et~al.
\newblock Real-time cell cycle imaging during melanoma growth, invasion, and
  drug response.
\newblock Pigment Cell {\&} Melanoma Research. 2014;27:764--776.

\bibitem{Davies2009}
Davies MA, Stemke-Hale K, Lin E, Tellez C, Deng W, Gopal YN, et~al.
\newblock Integrated molecular and clinical analysis of {AKT} activation in
  metastatic melanoma.
\newblock Clinical Cancer Research. 2009;15:7538--7546.

\bibitem{Hoek2006}
Hoek KS, Schlegel NC, Brafford P, Sucker A, Ugurel S, Kumar R, et~al.
\newblock Metastatic potential of melanomas defined by specific gene expression
  profiles with no {BRAF} signature.
\newblock Pigment Cell Research. 2006;19:290--302.

\bibitem{Smalley2007}
Smalley KSM, Contractor R, Haass NK, Kulp AN, Atilla-Gokcumen GE, Williams DS,
  et~al.
\newblock An organometallic protein kinase inhibitor pharmacologically
  activates {p53} and induces apoptosis in human melanoma cells.
\newblock Cancer Research. 2007;67:209--217.

\bibitem{Smalley2007a}
Smalley KSM, Contractor R, Haass NK, Lee JT, Nathanson KL, Medina CA, et~al.
\newblock {Ki}67 expression levels are a better marker of reduced melanoma
  growth following {MEK} inhibitor treatment than phospho-{ERK} levels.
\newblock British Journal of Cancer. 2007;96:445--449.

\bibitem{Spoerri2017}
Spoerri L, Beaumont KA, Anfosso A, Haass NK.
\newblock Real-time cell cycle imaging in a 3D cell culture model of melanoma.
\newblock Methods in Molecular Biology. 2017;1612:401--416.

\bibitem{Sakaue_Sawano2008}
Sakaue-Sawano A, Kurokawa H, Morimura T, Hanyu A, Hama H, Osawa H, et~al.
\newblock Visualizing spatiotemporal dynamics of multicellular cell-cycle
  progression.
\newblock Cell. 2008;132:487--498.

\bibitem{Allen2010}
Allen LJS.
\newblock An Introduction to Stochastic Processes with Applications to Biology.
\newblock 2nd ed. Taylor \& Francis Ltd.; 2010.

\bibitem{Weber2014}
Weber TS, Jaehnert I, Schichor C, Or-Guil M, Carneiro J.
\newblock Quantifying the length and variance of the eukaryotic cell cycle
  phases by a stochastic model and dual nucleoside pulse labelling.
\newblock {PLoS} Computational Biology. 2014;10:e1003616.

\bibitem{Yates2017}
Yates CA, Ford MJ, Mort RL.
\newblock A multi-stage representation of cell proliferation as a Markov
  process.
\newblock Bulletin of Mathematical Biology. 2017;79:2905--2928.

\bibitem{Vittadello2019}
Vittadello ST, McCue SW, Gunasingh G, Haass NK, Simpson MJ.
\newblock Mathematical models incorporating a multi-stage cell cycle replicate
  normally-hidden inherent synchronization in cell proliferation.
\newblock Journal of the Royal Society Interface. 2019;16:20190382.

\bibitem{Chao2019}
Chao HX, Fakhreddin RI, Shimerov HK, Kedziora KM, Kumar RJ, Perez J, et~al.
\newblock Evidence that the human cell cycle is a series of uncoupled,
  memoryless phases.
\newblock Molecular Systems Biology. 2019;15:e8604.

\bibitem{Gavagnin2019}
Gavagnin E, Ford MJ, Mort RL, Rogers T, Yates CA.
\newblock The invasion speed of cell migration models with realistic cell cycle
  time distributions.
\newblock Journal of Theoretical Biology. 2019;481:91--99.

\bibitem{MATLAB:lsqnonlin}
{MATLAB lsqnonlin}. {S}olve nonlinear least-squares (nonlinear data-fitting)
  problems ({R}2019b). {A}ccessed {F}ebruary 2020.; 2019.
\newblock Available from:
  \url{https://mathworks.com/help/optim/ug/lsqnonlin.html}.

\bibitem{Coleman1996}
Coleman TF, Li Y.
\newblock An interior trust region approach for nonlinear minimization subject
  to bounds.
\newblock {SIAM} Journal on Optimization. 1996;6:418--445.

\bibitem{Smith2011}
Smith H.
\newblock An Introduction to Delay Differential Equations with Applications to
  the Life Sciences.
\newblock Springer-Verlag GmbH; 2011.

\end{thebibliography}

\end{document}